\documentclass[lettersize,journal]{IEEEtran}
\usepackage{amsmath,amsfonts}
\usepackage{array,tabularx,makecell}
\usepackage{graphicx}
\usepackage{stfloats}                     
\usepackage{algorithm,algorithmic}
\makeatletter
\newcommand{\INPUT}{\item[\algorithmicinput]}
\newcommand{\OUTPUT}{\item[\algorithmicoutput]}
\newcommand{\algorithmicinput}{\textbf{Input:}}
\newcommand{\algorithmicoutput}{\textbf{Output:}}
\makeatother
\usepackage{cite}                         
\usepackage{textcomp}
\usepackage{verbatim}
\usepackage[caption=false,font=footnotesize]{subfig}
\usepackage{xcolor}
\usepackage{microtype}
\usepackage{xurl}
\usepackage[colorlinks=true,allcolors=blue]{hyperref}
\usepackage{orcidlink}
\begin{document}
\title{Detecting Fake Reviewer Groups in Dynamic Networks: An Adaptive Graph Learning Method}

\author{Jing Zhang\orcidlink{0000-0003-2494-8077}, Ke Huang\orcidlink{0009-0001-2253-464X}, Yao Zhang\orcidlink{0000-0002-1276-5399},~\IEEEmembership{Member,~IEEE}, Bin Guo,~\IEEEmembership{Senior Member,~IEEE}, and Zhiwen Yu,~\IEEEmembership{Senior Member,~IEEE}
\IEEEcompsocitemizethanks{
    \IEEEcompsocthanksitem This work was supported in part by the National Natural Science Fund for Distinguished Young Scholars (62025205), in part by the National Natural Science Foundation of China under Grant 62541327, Grant 62532009, Grant U25B2042, and Grant 62302396, in part by the Natural Science Foundation of Shaanxi Province (Grant No.2024JC-YBQN-0665). \emph{(Corresponding author: Yao Zhang)}
    \IEEEcompsocthanksitem Jing Zhang and Ke Huang are with the School of Computer Science and Technology, Xi’an University of Science and Technology, Xi’an 710600, China.
    \IEEEcompsocthanksitem Yao Zhang and Bin Guo are with the School of Computer Science, Northwestern Polytechnical University, Xi’an 710021, China.
    \IEEEcompsocthanksitem Z. Yu is with Harbin Engineering University, Harbin, Heilongjiang, China, and also with the School of Computer Science, Northwestern Polytechnical University, Xi'an, Shaanxi, China.
    }
}

\markboth{Journal of \LaTeX\ Class Files,~Vol.~14, No.~8, August~2021}%
{Shell \MakeLowercase{\textit{et al.}}: A Sample Article Using IEEEtran.cls for IEEE Journals}


\maketitle

\begin{abstract}
The proliferation of fake reviews, often produced by organized groups, undermines consumer trust and fair competition on online platforms. These groups employ sophisticated strategies that evade traditional detection methods, particularly in cold-start scenarios involving newly launched products with sparse data.
To address this, we propose the \underline{D}iversity- and \underline{S}imilarity-aware \underline{D}ynamic \underline{G}raph \underline{A}ttention-enhanced \underline{G}raph \underline{C}onvolutional \underline{N}etwork (DS-DGA-GCN), a new graph learning model for detecting fake reviewer groups. DS-DGA-GCN achieves robust detection since it focuses on the joint relationships among products, reviews, and reviewers by modeling product-review-reviewer networks. DS-DGA-GCN also achieves adaptive detection by integrating a Network Feature Scoring (NFS) system and a new dynamic graph attention mechanism. The NFS system quantifies network attributes, including neighbor diversity, network self-similarity, as a unified feature score. The dynamic graph attention mechanism improves the adaptability and computational efficiency by captures features related to temporal information, node importance, and global network structure.
Extensive experiments conducted on two real-world datasets derived from Amazon and Xiaohongshu demonstrate that DS-DGA-GCN significantly outperforms state-of-the-art baselines, achieving accuracies of up to \textbf{89.8\% and 88.3\%}, respectively.

\end{abstract}

\begin{IEEEkeywords}
Fake Review Detection; Graph Learning; NFS System; Dynamic Graph Attention Mechanism
\end{IEEEkeywords}

\section{Introduction}
\IEEEPARstart{I}{n}  online markets, detecting fake reviews is crucial for maintaining a fair and trustworthy environment \cite{1111111he2022market}. 
The scale of this issue is substantial; for instance, a recent study estimated that fraudulent reviews influence billions of dollars in online spending annually, with a significant portion of sellers admitting to using them to boost sales \footnote{ https://www.weforum.org/agenda/2021/11/fake-reviews-online-shopping-cost/}. 
According to a recent survey\footnote{https://help-center.pissedconsumer.com/online-review-trends-and-statistics/}, 90.6\% of consumers read online reviews before purchasing. While genuine positive reviews can build trust, excessive fake reviews distort consumer choices and undermine market fairness \cite{2222222222tadelis2016reputation}. To protect their reputation, platforms still rely heavily on manual screening, which consumes significant human resources.

Despite over a decade of research on fake review detection \cite{wu2020fake}, fake reviews remain widespread \cite{44444444mukherjee2012spotting,5555555he2022detecting}. 
For example, a significant proportion of Amazon product reviews are still fabricated\cite{1111111he2022market}, undermining consumer trust; according to the same survey, only 75.5\% of consumers continue to place confidence in online reviews\footnotemark[1].
Two challenges make fake review detection particularly difficult:

\begin{itemize}
    \item \emph{Organized fake review generation.} Fake reviewers often collaborate in groups organized via off-platform private channels, such as dedicated social media groups or messaging apps. This coordination is invisible to e-commerce platforms, making it difficult to detect collusion based on on-site data alone. They may deliberately misalign texts and ratings (e.g., negative text with high star rating) or coordinate to target the same products at different times to mimic organic review patterns and avoid burst detection \cite{xu2020detect,777777lim2010detecting}. Such tactics are designed to circumvent systems that focus only on content anomalies or individual accounts \cite{si2020shilling, mukherjee2013yelp}.

    \item \emph{Newly launched products.} In the early stage of a product launch, only a few reviews are available. Traditional methods that rely on abundant content or static structural features perform poorly here: sparse content limits behavioral signals, and the reviewer–product subgraph is extremely small and star-shaped, which undermines static heuristics. Fake groups can thus quickly dominate the initial review set, misleading both consumers and later detection algorithms\cite{new333mukherjee2012spotting}.
\end{itemize}

The existing research on the challenge of fake reviewer group detection can be categorized into two types: \emph{relationship-based} and \emph{network-based} detection methods.
The relationship-based detection methods focus on identifying the relationships between reviews and reviewers \cite{44444444mukherjee2012spotting, 888888olsson2024fakex}, as well as distinguishing connections between fake reviewers and genuine reviewers \cite{liu2018contrast}. For example, a fundamental principle is to analyze the features of reviews, such as ratings, votes, and text content, to identify whether reviews are posted by similar reviewers. This approach is effective since fake reviewers within a group usually exhibit similar patterns in their reviews' features. However, relationship-based methods are developed using historical review data. Human reviewers may evade the detection mechanism by writing fake reviews indistinguishable from organic reviews. The network-based detection methods aim to identify products that use fake reviews by exploiting network-related features extracted from product-review-reviewer networks \cite{5555555he2022detecting, 9999999akoglu2013opinion, A444444zhao2023rhgnn, yu2024temporal}.
The development of product-review-reviewer networks is based on an insightful finding revealed by He et al \cite{5555555he2022detecting} that the products buy fake reviews are highly clustered. The network-based detection methods exhibit stronger resistance to manipulation, as the network features are more difficult to manipulate or forge \cite{9999999akoglu2013opinion}.

Despite existing works can achieve decent performance in their targeted datasets, we lack adaptive methods to handle the aforementioned challenge caused by newly launched products. An Amazon Statistics report shows that there are thousands of newly-launched products every day\footnote{https://amzscout.net/blog/amazon-statistics/}. To address that, in this work, our goal is to enhance the adaptability of fake reviewer group detection in the real-world e-commence platforms that involve the launch of new products at various scales. We leverage the principle of graph learning technology to design our fake reviewer group detection algorithm. Researcher have devoted considerable effort to study fake reviewer group detection methods utilizing various graph learning technologies, including graph convolution \cite{A11111111fayazi2015uncovering, A22222wang2019fdgars}, graph pooling and clustering \cite{A33333wang2021decoupling}, as well as graph attention mechanism \cite{A444444zhao2023rhgnn}. These approaches effectively harness the capabilities of graph learning techniques in processing the complex topological data.
However, we note that the product-review-reviewer network for an e-commence platform is highly dynamic since numerous products will be launched or pulled every day. This phenomenon will incur performance degradation of benchmark-trained graph learning models as their training relies on the static graph assumption.

In this paper, we propose an adaptive graph learning-based fake reviewer group detection algorithm. In contrast to existing graph learning based approaches, our algorithm exploits the intrinsic features of the dynamic product-review-reviewer network, including the diversity of nodes and the self-similarity the network. To efficiently extract these features, our algorithm incorporates the dynamic graph attention mechanism into a graph convolutional network, forming a \underline{D}iversity- and \underline{S}imilarity-aware \underline{D}ynamic \underline{G}raph \underline{A}ttention-enhanced \underline{G}raph \underline{C}onvolutional \underline{N}etwork, dubbed as DS-DGA-GCN. Our algorithm is able to adaptively quantify the network-related features, accordingly enhancing the accuracy, efficiency, and robustness of detecting fake reviewer groups. Specifically, the design rationales of our algorithm are twofold:

\emph{1) How to measure the importance of dynamic attributes in the product-review-reviewer network?} Existing network-based detection methods for fake reviewer groups have shown that the node diversity and network self-similarity are key attributes in revealing potentially anomalous behaviors within a network. However, given dynamic relationships among products, reviews, and reviewers, the impact of these attributes on fake reviewer group detection is unknown. As aforementioned, adapting graph learning technologies to extract network-related features becomes viable, given static graph assumption. The core principle of designing our adaptive graph learning algorithm is the imperative need to ensure that the GCN model can identify the importance of attributes within a network. Our NFS module uses an innovative scoring mechanism to integrate diverse network attributes, transforming them into intuitive score representations. This lightweight approach is crucial for quantifying their collective impact in detecting fake reviewer groups.

\emph{2) How to adaptively utilize the importance of attributes in fake reviewer group detection?} The importance of attributes within the network fluctuates with changes in time and network structure. Graph attention mechanism has been proved as an efficient tool that enables nodes within the network aggregate neighborhoods' features based on feature importance \cite{velivckovic2017graph}. However, we lack an efficient graph attention mechanism that can handle the complex and dynamic nature of the product-review-reviewer network. We develop a dynamic graph attention mechanism that can be incorporated into GCNs. By integrating temporal information, node importance scores, and global network features, this mechanism dynamically learns the importance of neighboring nodes and prioritizes the extraction of key information during the aggregation process. As a result, it enhances the adaptability of fake reviewer group detection in complex and dynamic product-review-reviewer networks.

The main contributions of this paper are as follows:
\begin{itemize}
   \item {\emph{Conceptual Innovation:}} To the best of our knowledge, this is the first work that takes adaptability as the core optimization dimension of fake reviewer group detection, given dynamic product-review-reviewer network in realistic e-commerce platforms.
   \item {\emph{Algoritgm Design:}} The newly proposed algorithm, DS-DGA-GCN, combines the NFS system the GCN model embedded by a new dynamic graph attention mechanism, achieving significant improvements in detection accuracy and adaptability. The NFS system integrates various network attributes to provide high-quality input features while the dynamic graph attention mechanism achieves adaptive aggregation of temporal information among the global network.
   \item {\emph{Experimental Validation:}} On two real-world datasets, our method outperforms baselines in Accuracy, Recall, F1, and AUROC, and achieves more stable performance under both temporal shift and newly launched product sparsity.
\end{itemize}

The structure of this paper is as follows. Section II discusses related works in the field. Section III details the proposed  DS-DGA-GCN algorithm. Section IV presents the experimental settings, performance analysis, and experimental results while Section V concludes the paper with a summary of the main contributions and an outlook on future research directions.

\section{Related work}

{
\subsection{Content and Behavioral-based Detection Methods}

Content- and behavioral-based approaches leverage textual semantics and user interaction signals to identify fake reviews. The central idea is to mine informative cues from review texts, ratings, and behavioral traces so as to discriminate organic vs.\ fraudulent activity.

Xu et al.\ proposed a multimodal learning model ``MMD'' that integrates rating signals with review content; via metric learning it effectively identifies Professional Malicious Users (PMUs) and boosts detection performance \cite{xiu1xu2020detect}. Li et al.\ introduced the notion of reviewer groups and modeled collusion among reviewers, thereby improving the identification of professional fake reviewers \cite{li2021detectionAA12}. Beyond reviews, Li et al.\ studied large-scale fake clicks in recommendation systems and formulated the ``Ride Item's Coattails'' attack that forges clicks to link low-quality with popular items; their detection techniques analyze click patterns to harden systems against such attacks \cite{xiu2li2021large}. Yang et al.\ mined fraudsters in mobile social networks using semi-supervised and metric learning, obtaining precise characterizations of fraudulent behaviors in dynamic environments \cite{xiu3yang2019mining}. Together, these works underscore the value of content and behavioral modeling for large-scale fraud analysis in complex interaction scenarios.
Content and behavioral-based approaches are effective for detecting local anomalies in texts or individual user activity. 
However, they are limited in capturing (i) cross-entity structural patterns across products, reviews, and reviewers, and (ii) temporal coordination such as bursty or staged campaigns. 
DS-DGA-GCN explicitly models a dynamic heterogeneous graph, where NFS encodes structural signals (e.g., neighborhood diversity and self-similarity) and the dynamic attention mechanism propagates information over time. 
This design links dispersed reviewers through typed relations and temporal dynamics, enabling the detection of coordinated behaviors beyond the reach of content- or behavior-centric methods.

\subsection{Collaborative Group Detection in Networks}

A complementary line models collaboration explicitly on reviewer graphs. Wei et al.\ analyzed reviewer behaviors and inter-node structures, defining node diversity and self-similarity; with cumulative distribution functions and two-hop subgraphs they identified concealed fraudulent clusters at scale \cite{jinrui2023massive99}. Soni et al.\ designed a top-down framework that combines deep random walks with improved semi-supervised clustering and background knowledge, achieving strong accuracy on Google Play data \cite{soni2018effective1100}. Es-Sabery et al.\ applied social network analysis and community detection, using weighted Gini coefficients to measure group purity and uncover coordinated fake review groups \cite{gai99es2021mapreduce}. These works highlight the utility of network behavior modeling and community detection in exposing collusion. 

Cao et al.\ further proposed an unsupervised end-to-end method that uses a modular GCN to discover tightly-knit groups and integrates individual- and group-level anomaly indicators into a unified suspiciousness score \cite{cao2021fakeAA13}. Unsupervised formulations improve scalability and portability across platforms by reducing reliance on labels.
Classical group-detection methods often rely on handcrafted heuristics (e.g., purity scores, Gini indices, subgraph thresholds) and static snapshots. 
Such heuristics are brittle under evolving adversarial strategies and sensitive to parameter choices. 
DS-DGA-GCN replaces these heuristics with learnable NFS that integrates multiple structural cues, and performs temporal message passing on dynamic graphs. 
This reduces threshold sensitivity, captures coordination across time, and remains scalable to large e-commerce graphs through pooling and sampling.

\subsection{Graph Learning-based Detection Methods}

Graph learning methods model user–item interactions as graphs and learn to capture higher-order structures indicative of fraud. Techniques span GCNs and broader GNN families that propagate information over the topology to surface anomalous substructures.

Fan et al.\ introduced Graph Decomposition Network (GDN) to address Structural Distribution Shift (SDS) in graph anomaly detection; by frequency-domain decomposition it enhances normal-node homogeneity and mitigates anomalous heterogeneity, improving robustness on dynamic networks \cite{xiu4fan2023adversarial}. Gao et al.\ revisited SDS and proposed an enhanced GDN that resists high heterophily for anomalies while reinforcing homophily for normal nodes, improving generalization under adversarial conditions \cite{xiu10gao2024revisiting}. Wang et al.\ proposed Deep Cluster Infomax (DCI), a self-supervised scheme that decouples representation learning from classification and partitions graphs into subgraphs to capture intrinsic properties, yielding gains in semi-supervised settings \cite{xiu8wang2021decoupling}.

Topology-centric approaches also remain effective. Fayazi et al.\ used a Markov Random Field-based clustering over examiner–examiner graphs to uncover crowdsourced manipulation, outperforming traditional and SimRank-based detectors \cite{xiu6fayazi2015uncovering}. Liu et al.\ proposed HoloScope, a contrastive metric that fuses graph topology with burstiness signals to emphasize fraud–normal contrasts in rich graphs \cite{xiu9liu2018contrast}. Hooi et al.\ developed FRAUDAR to detect fraud in bipartite user–product graphs, remaining robust to camouflage and hijacked accounts \cite{xiu5hooi2016fraudar}. 

Within review ecosystems, Wang et al.\ presented FdGars, the first GCN-based detector at scale for app review systems, constructing user–app graphs to extract structural evidence and achieve high accuracy \cite{xiu7wang2019fdgars}. Yu et al.\ proposed GFDN with a Temporal Group Dynamics Analyzer (TGDA) to model temporal aspects of group fraud in a semi-supervised framework \cite{xiu11yu2024temporal}. Overall, GNN-based methods excel at capturing intricate network relationships and temporal dynamics for large-scale fraud detection.
In particular, heterogeneous GNNs and temporal GNNs are directly pertinent to our setting. HetGNN~\cite{new11zhang2019heterogeneous} and TGN~\cite{new12rossi2020temporal} represent two state-of-the-art approaches for handling the heterogeneous and dynamic aspects of our data, respectively, and thus serve as strong baselines. HetGNN excels at capturing type-specific relations but lacks explicit temporal modeling, whereas TGN preserves temporal states but does not natively exploit type-aware structures.
Our proposed DS-DGA-GCN is designed to unify these two perspectives through a synergistic architecture. It first distills complex structural properties into a robust NFS and then incorporates this score, alongside temporal information, into its dynamic attention mechanism during aggregation. 
This unique integration is crucial for enhancing model robustness, particularly under conditions of review data sparsity and in cold-start scenarios-a key advantage that we empirically validate in our experiments on newly-launched products.

}

\begin{figure*}[!t]
\centering
\includegraphics[width=\linewidth]{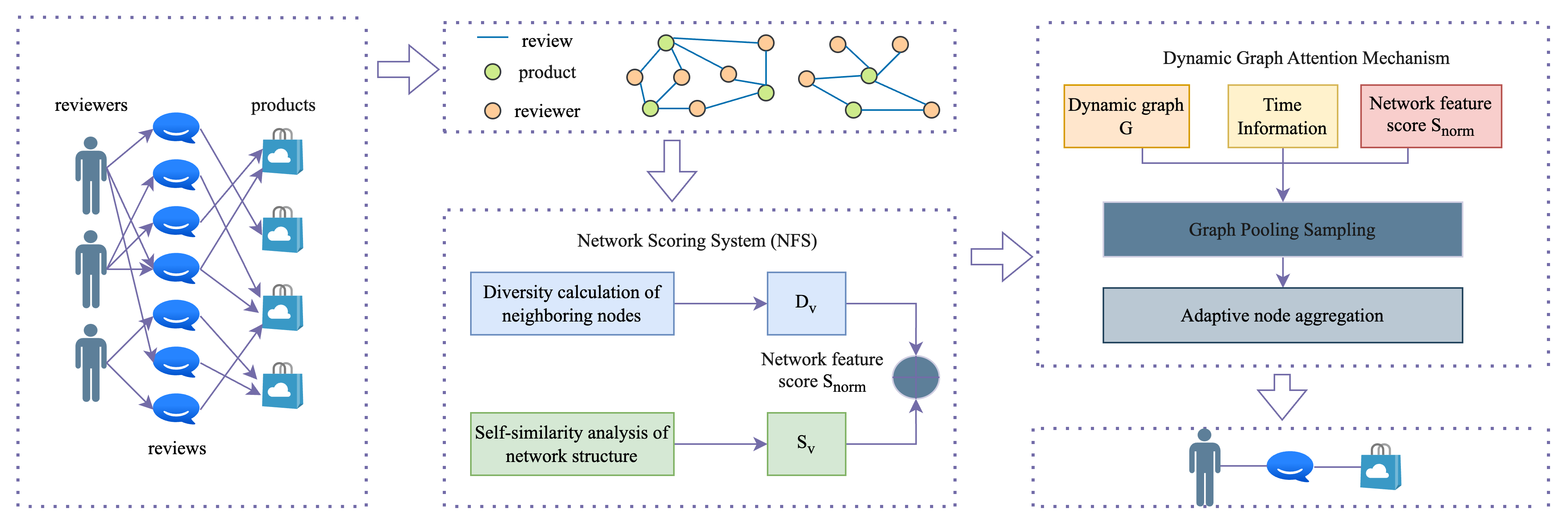}
\caption{DS-DGA-GCN Model Framework Diagram. Left: consumers, reviewers, products; Middle: where is $C_v$ and $T_v$}
\label{fig_2}
\end{figure*}

\section{Fake Review Group Detection Algorithm Based on DS-DGA-GCN}
In this section, we model the product-review-reviewer network as a dynamic heterogeneous graph in order to capture the relationships of products, reviews, and reviewers. 
The following definitions describe the main components within the network:

\begin{itemize}
   \item {\emph{Products:}} Denoted as $P=\{p_1,p_2,...,p_N\}$, representing the set of $N$ products on the platform.
   \item {\emph{Reviewers:}} Denoted as $U=\{u_1,u_2,...,u_M\}$, representing the set of $M$ potential reviewers, including fake reviewers and genuine consumers, who can post reviews for products.
   \item {\emph{Reviews:}} Denoted as $R=\{r_1,r_2,...,r_L\}$, representing the set of $L$ reviews written by reviewers.
\end{itemize}

Each review $r_k$, characterized by its attributes including timestamp $t_k$, content $c_k$, and rating $s_k$, is associated with a reviewer $u_{i(k)} \in U$ and a product $p_{j(k)} \in P$. 
The dynamic heterogeneous graph is defined as $G=(V,E)$. $V$ refers to the set of nodes where each node is a reviewer $u_i \in U$ or a product $p_j \in P$. $E$ refers to the set of edges where $e_k=(u_{i(k)},p_{j(k)})$ denotes the attributes of the review $r_k$ written by the reviewer $u_{i(k)}$ for the product $p_{j(k)}$. {The edge consists of a tuple, including $t_k$, $c_k$, and $s_k$.}

The graph $G$ evolves at discrete time steps as new reviews are added. At each time step $t$, the graph is denoted as $G_t=(V,E_t)$, where $E_t$ includes all edges up to time $t$.

We then define the core relationships within the graph as follows:
\begin{itemize}
   \item {\emph{Reviewer Feature Matrix:}} We define the reviewer feature matrix as $H^u \in \mathbb{R} ^{N \times d_u}$, where each row corresponds to the features of a reviewer node. 
   \item {\emph{Product Feature Matrix:}} We define the product feature matrix as $H^p \in \mathbb{R} ^{N \times d_p}$, where each row corresponds to the features of a product node.
   \item {\emph{Adjacency Matrix:}} At time $t$, the relationship between reviewers and products is represented by the adjacency matrix $A_t \in \mathbb{R} ^{M \times N}$, where the element $A_t(i,j)=0$ indicates that reviewer $u_i$ has reviewed product $p_j$ before time $t$, otherwise the element is $0$. 
\end{itemize}


The proposed DS-DGA-GCN encompasses two modules to process the graph $G_t$, including an NFS and a dynamic graph attention mechanism. Specifically, the NFS module quantifies {node diversity and network self-similarity as a network feature score to provide node importance information for the model}. By combining the network feature scoring process and the dynamic graph attention mechanism, the DS-DGA-GCN, as shown in Fig. \ref{fig_2}, is capable of detecting fake reviewer groups in complex dynamic product-review-reviewer networks. The dynamic graph attention mechanism is consist of two submodules, including graph pooling and sampling, as well as adaptive node aggregation. {The former divides the graph into key subgraphs to mitigate the impact of redundant or unimportant nodes and edges within the network and thus reduce computational complexity. The latter dynamically learns the relationship weights among nodes using the attention mechanism to efficiently aggregate node information.}

\subsection{NFS}
\subsubsection{Algorithm Overview}
To quantify the potential anomalousness of nodes, our model integrates multiple network features into a unified score. Specifically, the two primary features used are:
\begin{itemize}
    \item {\emph{Neighbor Diversity ($D_v$):}} Assesses the diversity of a node's neighbors. For reviewer nodes, lower diversity may indicate interactions mainly with specific products, potentially signaling coordinated behavior \cite{song2005selfBB24}.
    \item {\emph{Network Self-Similarity ($S_v$):}} Measures the self-similarity of a node's ego-network structure. High self-similarity can indicate repetitive interaction patterns, characteristic of collusive activities \cite{gallos2007reviewBB25}.
\end{itemize}

The workflow for calculating the final anomaly score is presented in Algorithm \ref{alg:nfs_calculation}. The process begins with the pre-computed diversity $D_v$ and self-similarity $S_v$ scores for all nodes. These two features are concatenated to form an initial feature vector for each node. This vector then undergoes preprocessing (standardization and PCA). A pre-trained linear SVM provides a weight vector, $\mathbf{w}$, which projects the processed features into a raw anomaly score. Finally, these raw scores are normalized to produce the final anomaly score $S_{\text{norm},v}$, for each node.

\begin{algorithm}[t]
\caption{NFS Score Calculation}
\label{alg:nfs_calculation}
\begin{algorithmic}[1]
\STATE \textbf{Input:} 
\STATE \quad $D$: Array of pre-computed diversity scores for all nodes $v \in V$.
\STATE \quad $S$: Array of pre-computed self-similarity scores for all nodes $v \in V$.
\STATE \quad $\text{PCA}_{\text{model}}$: A pre-trained PCA transformation model.
\STATE \quad $\mathbf{w}$: A pre-trained SVM weight vector.

\STATE \textbf{Output:} 
\STATE \quad $S_{\text{norm}}$: Array of final normalized anomaly scores for all nodes $v \in V$.

\STATE \textbf{Initialize} feature matrix $\mathbf{F}$ of size $|V| \times 2$.

\FOR{each node $v \in V$}
    \STATE $\mathbf{F}[v] \leftarrow [D[v], S[v]]$ \COMMENT{Concatenate features for each node.}
\ENDFOR

\STATE $\mathbf{F}_{\text{std}} \leftarrow \text{Standardize}(\mathbf{F})$ \COMMENT{Standardize features (zero mean, unit variance).}
\STATE $\mathbf{Z} \leftarrow \text{PCA}_{\text{model}}.\text{transform}(\mathbf{F}_{\text{std}})$ \COMMENT{Apply PCA for dimensionality reduction.}

\STATE \textbf{Initialize} raw score array $S_{\text{nfs}}$ of size $|V|$.
\STATE $S_{\text{nfs}} \leftarrow \mathbf{Z} \cdot \mathbf{w}$ \COMMENT{Compute raw anomaly scores using SVM weights.}

\STATE $S_{\text{norm}} \leftarrow \text{MinMaxNormalize}(S_{\text{nfs}})$ \COMMENT{Normalize raw scores to range [0, 1].}

\STATE \textbf{return} $S_{\text{norm}}$
\end{algorithmic}
\end{algorithm}

\subsubsection{Diversity of Adjacent Nodes}
\begin{figure}[!t]
\centering
\includegraphics[width=\linewidth]{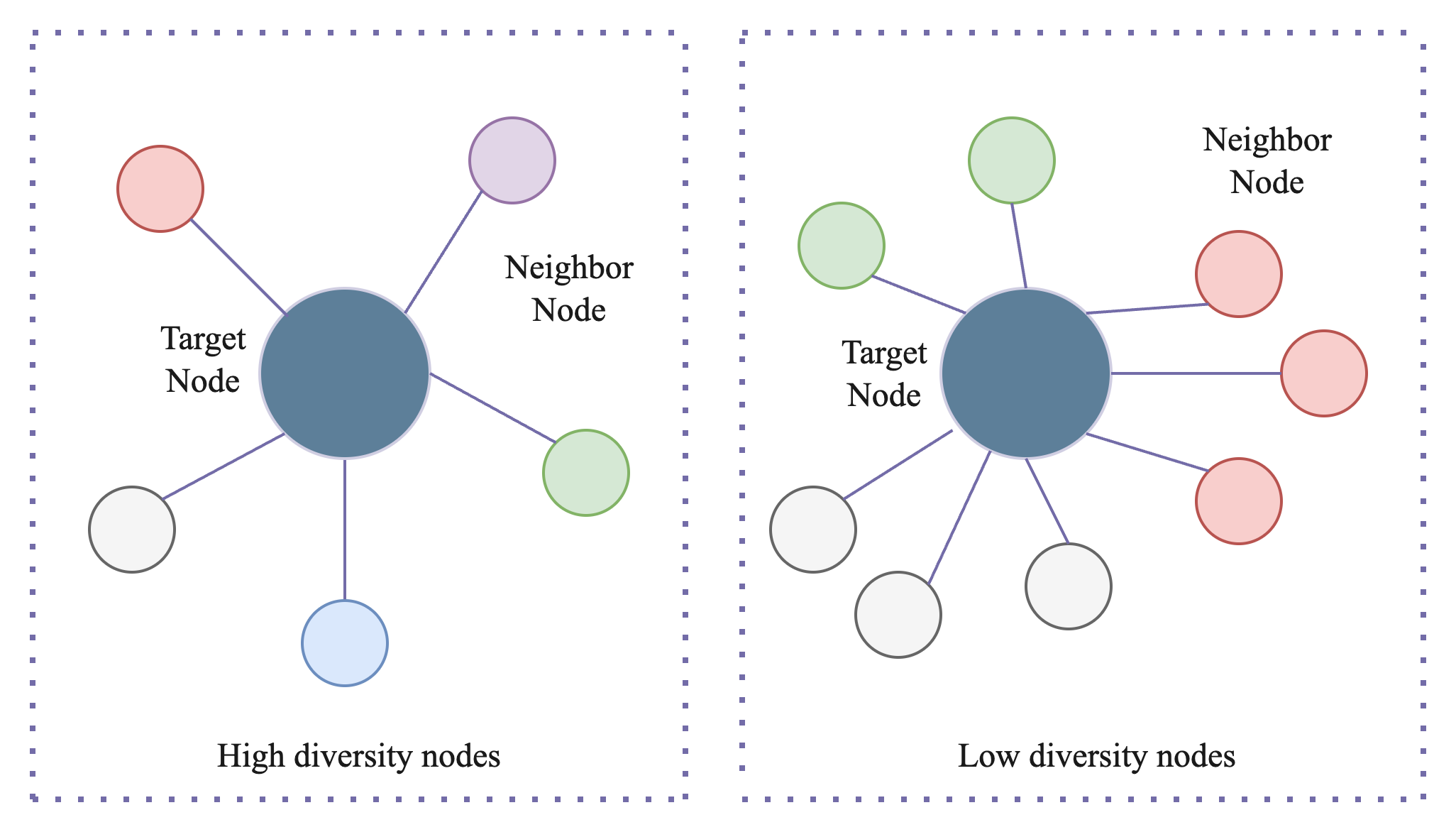}
\caption{Diversity of Adjacent Nodes.}
\label{fig_3}
\end{figure}

The diversity of nodes is a key signal for detecting anomalous behaviors in reviewer–product networks \cite{wang2012recommendationCC32}. 
For a specific product, diversity reflects the randomness of the behaviors of its reviewers. 
Fig.~\ref{fig_3} illustrates two cases: the left panel shows high diversity while the right panel shows low diversity, with reviewer types indicated by colors. 

We measure diversity using information entropy:
\begin{equation}
H(v)=-\sum_{k} P_k \log(P_k),
\end{equation}
where $P_k$ is the probability density of the centrality of the $k$-th reviewer.

Degree centrality captures the immediate scale of a node’s neighborhood.  
For a product connected to many one-time reviewers, the degree histogram becomes narrow and star-shaped, yielding low entropy—typical of coordinated spam campaigns:  
\begin{equation}
C(v)=\frac{deg(v)}{N-1},
\end{equation}
where $deg(v)$ is the degree of node $v$ and $N$ is the total number of nodes.

PageRank measures the authority of a node under random walks \cite{gai21page1998pagerank}.  
Reviewers confined to small collusive rings have low PageRank, while those engaged broadly across products and time obtain higher values, indicating organic activity:
\begin{equation}
PR(v)=(1-d)+d\sum_{u \in N(v)} \frac{PR(u)}{\kappa(u)},
\end{equation}
where $d$ is the damping factor (set to 0.85), and 
\begin{equation}
\kappa(u)=\sum_{v\in N(u)}A(u,v)
\end{equation}
denotes the degree of neighbor $u$ based on adjacency $A(u,v)$.

Let $k\in \mathcal{K}$ be the set of neighbor categories (e.g., reviewer types).  
For a target node $v$, denote the subset of neighbors in category $k$ as  
$U_k(v)=\{\,u\in N(v): u \text{ belongs to category } k\,\}$.  
We define the category proportion  
$p_k(v)=|U_k(v)|/|N(v)|$  
and the aggregated PageRank mass  
$PR_k(v)=\sum_{u\in U_k(v)} PR(u)$.  

By combining the local proportion and the global PageRank, the adjusted neighbor diversity weight is defined as:
\begin{equation}
\omega_k(v)=p_k(v)\cdot \frac{PR_k(v)}{\sum_{j\in\mathcal{K}} PR_j(v)}.
\end{equation}

\noindent
This design ensures that local neighbor diversity is weighted by global authority, so that concentrated low-quality neighbors cannot artificially inflate diversity scores.

In short, degree centrality captures local bursty patterns such as star-shaped spam clusters, 
while PageRank captures global authority and distinguishes collusive rings from widely connected reviewers. 
By combining them, our method balances sensitivity to local anomalies with robustness against pseudo-diversity created by low-authority spam rings.

Finally, we compute the diversity-adjusted entropy:
\begin{equation}
H_{pageRank}(v)=-\sum_k \omega_k(v)\log(\omega_k(v)),
\end{equation}
and normalize it as
\begin{equation}
\eta=\frac{H_{pageRank}(v)}{\max(H_{pageRank})}.
\end{equation}

\subsubsection{Self-Similarity of Network Structure}
\begin{figure}[!t]
\centering
\includegraphics[width=\linewidth]{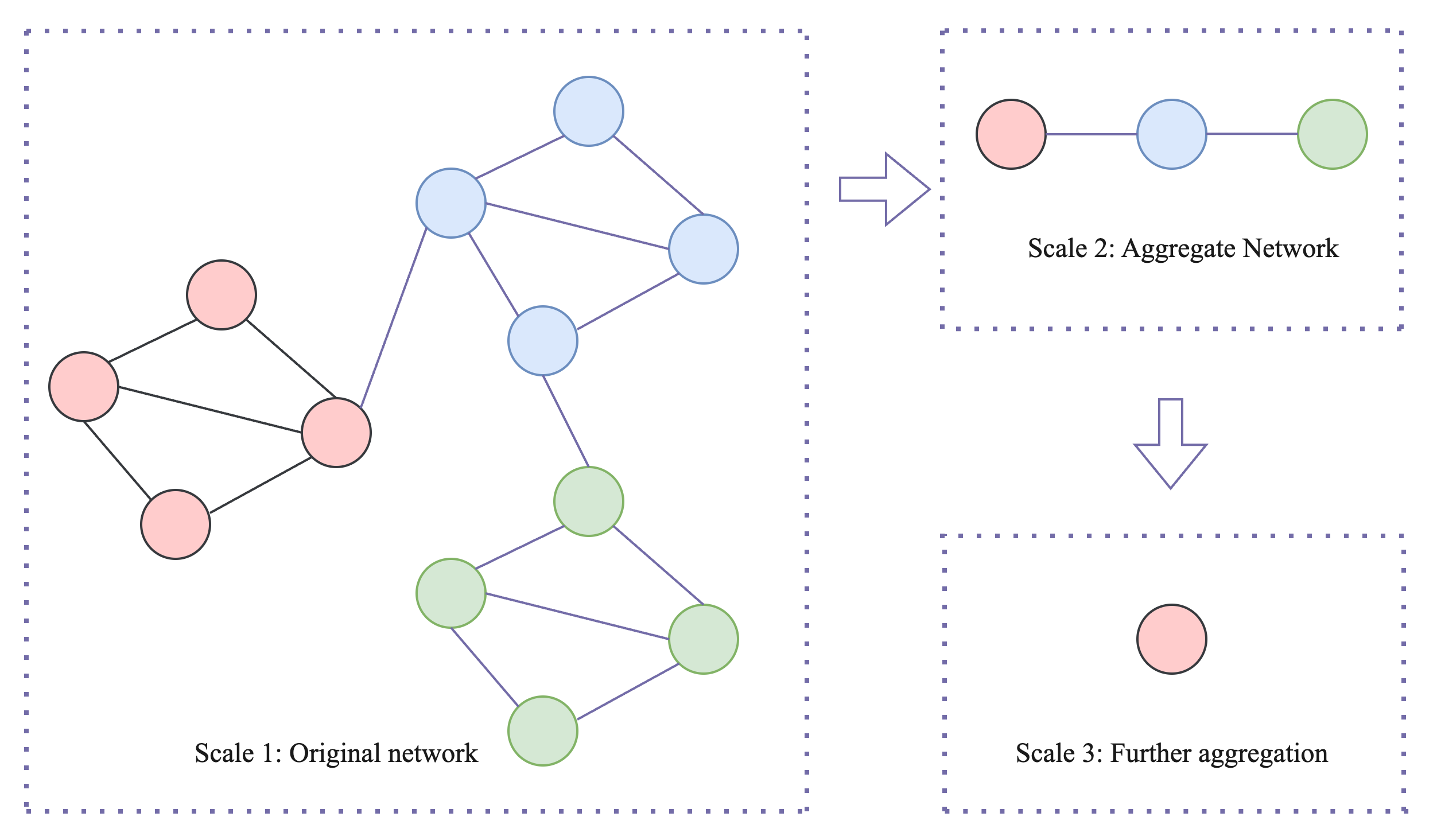}
\caption{Changes in Network Structure at Different Scales.}
\label{fig_4}
\end{figure}

Self-similarity is a key indicator of repetitive interaction patterns, often found in collusive activities within fake reviewer groups\cite{song2005selfBB29}. To effectively quantify this, we propose a composite self-similarity score, $S_v$, that integrates both geometric and spectral properties of the network, ensuring sensitivity to diverse fraudulent patterns like bursty stars and collusive rings. Fig.~\ref{fig_4} illustrates how network structures can exhibit such self-similar characteristics across different scales.

Our approach is built upon two complementary metrics. First, we compute the fractal dimension $C_f$, using the Box-Counting method, which captures geometric scaling and is sensitive to star-like structures. It is derived from the power-law $N(\epsilon) \propto (1/\epsilon)^{C_f}$, where $N(\epsilon)$ is the number of boxes of size $\epsilon$ to cover the network. Second, we analyze the network's spectral properties through the spectral exponent $\beta$, which characterizes the power-law distribution of the Laplacian eigenvalues, $P(\lambda) \sim \lambda^{-\beta}$. This metric is effective at identifying ring structures common in collusive groups.

Instead of using these metrics in isolation, we synthesize them into a single, robust score. The final integrated self-similarity score $S_v$ is defined as:
\begin{equation}
\label{eq:sv_score_final}
S_v = \big(\alpha S_{g,v} + (1-\alpha)S_{s,v}\big)\cdot M_v
\end{equation}
Here, $S_{g,v}$ and $S_{s,v}$ are the scores derived from the geometric $C_f$ and spectral $\beta$ analyses respectively. Each score is normalized and weighted by the quality-of-fit (e.g., R-squared value) of its power-law estimation to ensure reliability. The term $M_v$ is a multi-scale consistency factor, obtained via network coarse-graining\cite{palla2008fundamentalCC34}, which modulates the final score based on the stability of these patterns across scales. The hyperparameter $\alpha$ (default 0.5) balances the geometric and spectral contributions. This composite score $S_v$ serves as a key feature for our detection framework.

\subsubsection{Construction of NFS}
To effectively integrate the various structural and semantic features for anomaly detection, we construct a unified NFS\cite{gai23akoglu2015graph}. The process systematically transforms a node's high-dimensional features into a final, calibrated anomaly score.

First, for each node $v$, we define its initial feature vector $\mathbf{f}_v$ by concatenating its degree centrality $D_v$ and its self-similarity score $S_v$:
\begin{equation}
\label{eq:nfs_features}
\mathbf{f}_v = [D_v, S_v]
\end{equation}
This feature vector undergoes a two-step preprocessing: it is first standardized to have zero mean and unit variance, and then PCA is applied for dimensionality reduction, resulting in a processed feature vector $\mathbf{z}_v$.

Next, we learn a discriminative weight vector $\mathbf{w}$ using a linear SVM with a hinge-loss objective on the training data. This weight vector is used to project the processed features into a raw anomaly score $S_{\text{nfs},v}$:
\begin{equation}
\label{eq:nfs_raw_score}
S_{\text{nfs},v} = \mathbf{w}^\top \mathbf{z}_v
\end{equation}
This raw score is then normalized to the range $[0, 1]$ to produce the final node anomaly score $S_{\text{norm},v}$:
\begin{equation}
\label{eq:nfs_norm_score}
S_{\text{norm},v} = \frac{S_{\text{nfs},v} - \min(S_{\text{nfs}})}{\max(S_{\text{nfs}}) - \min(S_{\text{nfs}})}
\end{equation}

Finally, nodes are classified based on this score using an optimal threshold $t^*$. A node is labeled as suspicious if its score exceeds this threshold:
\begin{equation}
\label{eq:nfs_classification}
\text{Label}_v =
\begin{cases}
\text{Suspicious} & \text{if } S_{\text{norm},v} \geq t^* \\
\text{Normal} & \text{otherwise}
\end{cases}
\end{equation}
The threshold $t^*$ is not a fixed hyperparameter but is determined empirically after training. Specifically, we use the trained model to predict scores on a held-out validation set. The optimal threshold $t^*$ is then selected using \cite{new13youden1950index}, which finds the point that maximizes the difference between the true positive rate (TPR) and the false positive rate (FPR): $t^* = \arg\max_{t}(\text{TPR}(t) - \text{FPR}(t))$. This principled approach ensures that the threshold optimally balances sensitivity and specificity. Once determined, $t^*$ is fixed for evaluation on the test set.

\subsection{Dynamic Graph Attention Mechanism}
\subsubsection{Model Overview}
Based on NFS, this paper introduces a dynamic graph attention mechanism as the core innovative module of the DS-DGA-GCN model, aimed at efficiently detecting fake reviewer groups by leveraging temporal information, node importance, and global network structure.

\begin{figure}[!t]
\centering
\includegraphics[width=\linewidth]{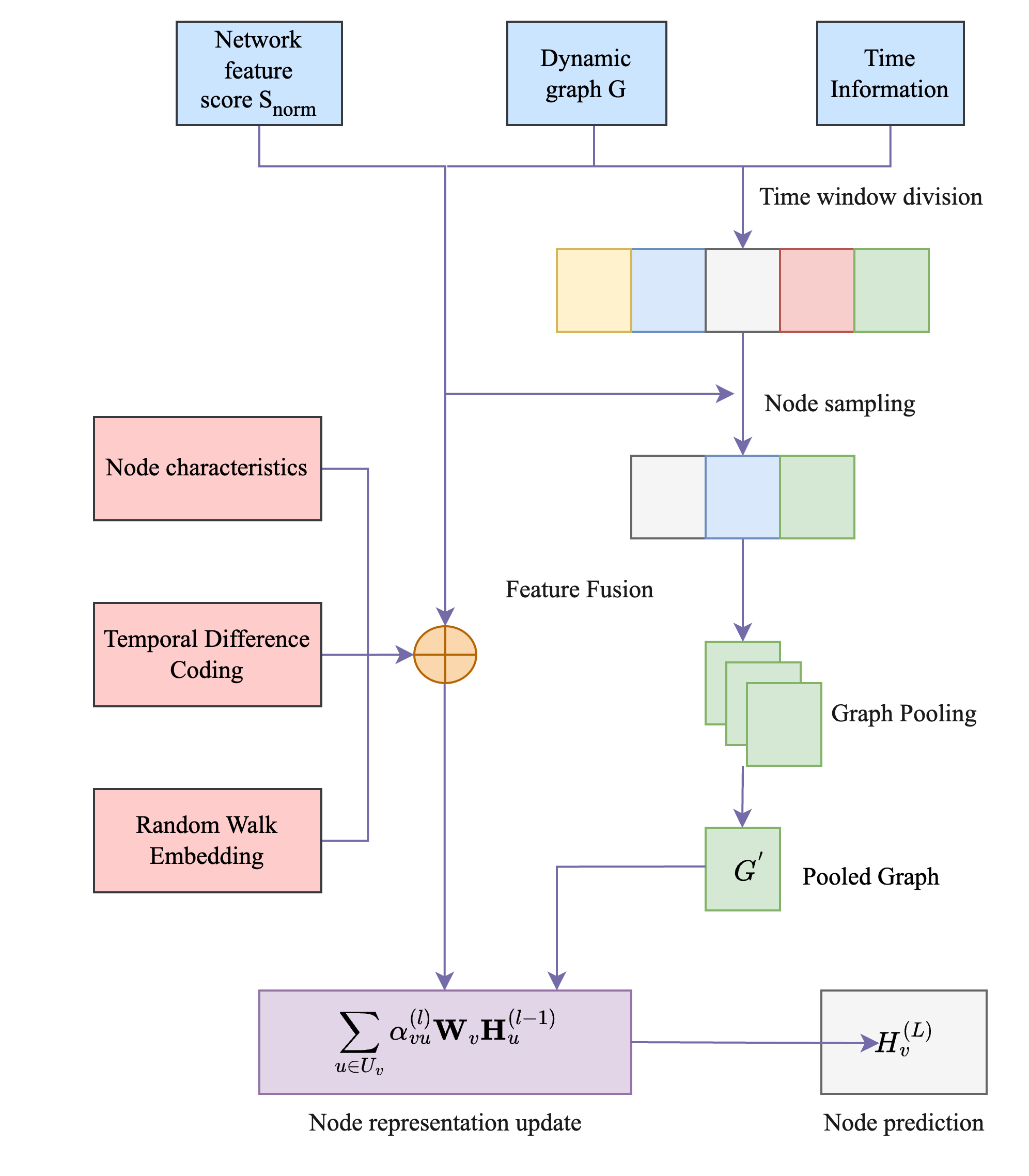}
\caption{Dynamic Graph Attention Mechanism.}
\label{fig_5}
\end{figure}

Traditional Graph Convolutional Networks (GCNs) are designed for static graphs and are unable to capture the time-varying characteristics of nodes and edges in dynamic networks. To address this, this paper proposes a dynamic graph attention mechanism framework by combining temporal information, node importance scores, and global structural information. This framework shown in Fig.\ref{fig_5} consists of the following three key components:

\begin{itemize}
   \item {\emph{Graph Pooling and Sampling:}} The dynamic graph is pooled and sampled by dividing it into time windows and evaluating node importance, resulting in simplified subgraphs that retain key network structural information while reducing computational complexity.
   \item {\emph{Adaptive Node Aggregation:}} On the simplified subgraph, the attention mechanism is used to dynamically learn the weights of neighboring nodes, integrating temporal differences, network feature scores, and global structural embeddings to capture complex relationships between nodes. 
   \item {\emph{Dynamic Graph Attention Updating:}} Through a multi-layer dynamic graph attention network, node representations are iteratively updated, mining the dynamic evolution features of nodes and their importance differences.
\end{itemize}



\subsubsection{Graph Pooling and Sampling}

When processing large-scale dynamic product-review-reviewer networks, the number of nodes and edges may grow rapidly over time, leading to high consumption of computational resources and difficulties in model training.
Additionally, the network contains a large number of redundant or outdated nodes and edges, which may interfere with the model's learning of important patterns. Therefore, how to effectively simplify and sample the dynamic graph while retaining the network's key structures and informative information becomes crucial. To improve computational efficiency and robustness of fake reviewer group detection, we develop a graph pooling and sampling method by resorting to \cite{tuchihuaying2018hierarchical}.
This method can capture and provide simplified and informative input for subsequent adaptive node aggregation, which is completed by selecting and merging key nodes and edges of the dynamic graph. 

Specifically, the method performs as follows. The dynamic graph is first divided into several time windows along the time series to capture the network's temporal evolution. Let the time series be $T=\{t_1,t_2,...,t_T\}$, and the entire time range is divided into $M$ time windows $\{\triangle t_1,\triangle t_2,...,\triangle t_M\}$, where the $m$-th time window is defined as:
\begin{equation}
\label{deqn_ex1a}
\triangle t_m=[t_{m-1},t_m).
\end{equation}

By integrating NFS, node degree $d_v$ and clustering coefficient $C_v$, the importance score of the node $v$ is defined as:
\begin{equation}
\label{deqn_ex1a}
I_v = \alpha_1 S_{\text{norm},v} + \alpha_2 \frac{d_v}{max_{u\in V}d_u} + \alpha_3 C_v,
\end{equation}
where $\alpha_1$, $\alpha_2$ and $\alpha_3$ are the weight coefficients subject to $\alpha_1+\alpha_2+\alpha_3=1$.

In each time window $\triangle t_M$, nodes are sampled by comparing their importance scores $I_v$ with the threshold $\theta$: 
\begin{equation}
\label{deqn_ex1a}
V_{\triangle t_m}^{'}=\{v\in V_{\triangle t_m}|I_v\geq \theta \}.
\end{equation}

For the sampled node set $V_{\triangle t_m}^{'}$, the corresponding edge set $\triangle t_m$ associated with these nodes in the time window $E_{\triangle t_m}^{'}$ is selected:

\begin{equation}
\label{deqn_ex1a}
E_{\triangle t_m}^{'}=\{e_{uv}\in E_{\triangle t_m}|u, v \in V_{\triangle t_m}^{'} \}.
\end{equation}

To preserve more structural information, the edge weights can be redefined. The importance of the edges is defined as $I_{e_{uv}}$:
\begin{equation}
\label{deqn_ex1a}
I_{e_{uv}}=\beta_1W_{uv}+\beta_2S_{uv},
\end{equation}
where $S_{uv}$ is the cosine similarity between nodes $u$ and $v$:
\begin{equation}
\label{deqn_ex1a}
S_{uv}=\frac{h_u^{\top}h_v}{||h_u||||h_v||},
\end{equation}
where $W_{uv}$ is the original weight of the edge, and $\beta_1$, $\beta_2$ are weight coefficients, satisfying $\beta_1+\beta_2=1$.

{The subgraph obtained through sampling is then pooled, with similar nodes and edges merged to further simplify the graph structure.} Clustering is performed on $V_{\triangle t_m}^{'}$, resulting in $K$ clusters $\{C_k\}$. The node representations are updated as follows:

\begin{equation}
\label{deqn_ex1a}
h_{C_k}=\frac{\sum_{v\in C_k}I_vh_v}{\sum_{v\in C_k}I_v}.
\end{equation}

Edge weight updates as:

\begin{equation}
\label{deqn_ex1a}
W_{c_kc_l}=\sum_{u\in c_k}\sum_{v \in c_l}I_{e_{uv}}.
\end{equation}

All pooled subgraphs from the time windows $\{\triangle t_m\}$ are merged to obtain the final simplified graph $G^{'}=(V^{'}, E^{'})$.

\subsubsection{Adaptive Node Aggregation}
Existing node aggregation methods include average aggregation, maximum aggregation, weighted aggregation, adaptive aggregation, and content aggregation. Average aggregation is the simplest form of node information integration, where for each node $V$, its new representation is the average of its neighboring nodes' representations. Its advantage lies in its simplicity, making it suitable for static networks, but it fails to distinguish the importance of neighboring nodes and ignores node heterogeneity. Maximum aggregation takes the maximum value from each dimension of the neighboring node representations, emphasizing the most significant features, but it may lose other important information and is not suitable for capturing the overall trend. Weighted aggregation assigns a preset weight to each neighboring node, considering the relationship strength between nodes; however, the weights need to be predefined and cannot be adjusted adaptively. Adaptive aggregation utilizes attention mechanisms to learn the weights of neighboring nodes based on node features. A typical example is the Graph Attention Network (GAT), which can adaptively learn the weights of neighboring nodes and capture node heterogeneity, but it does not consider temporal information and dynamic features, limiting its performance in dynamic graphs. Content aggregation typically refers to aggregating the feature representations of neighboring nodes. However, in some applications, edge features or combined node and edge features may also be aggregated to obtain richer information\cite{zhou2022networkCC38}\cite{kelley2017generalCC37}\cite{gai38yu2017spatio}.

In dynamic product-review-reviewer networks, the attributes of nodes and edges change over time, presenting the following challenges:

\begin{itemize}
   \item {\emph{Temporal Dynamics:}} The features of nodes and edges exhibit temporal dependencies, and traditional aggregation methods fail to capture temporal information.
   \item {\emph{Differences in Node Importance:}} Different reviewers and products have varying influences within the network, requiring differentiated treatment.
   \item {\emph{Complex Interactions:}} Fake reviewers may hide their behaviors through complex interaction patterns, making it difficult for simple aggregation methods to identify such behaviors.
\end{itemize}

Therefore, existing aggregation methods have limitations in the context of this paper and need to be improved to adapt to dynamic networks and node importance differences. To address these issues, this paper proposes a new adaptive node aggregation mechanism.


To capture the temporal relationships between nodes, this paper introduces temporal differences into the attention mechanism:
\begin{equation}
\label{deqn_ex1a}
\alpha_{uv}=\frac{exp(\sigma(\alpha^{\top}[W_qh_v||W_kh_u||W_t(t_v-t_u)]))}{\sum_{k\in N(v)}{exp(\sigma(\alpha^{\top}[W_qh_v||W_kh_k||W_t(t_v-t_k)]))}},
\end{equation}
where $t_v$ and $t_u$ are the timestamps of nodes $v$ and $u$, respectively. $W_q$, $W_k$ and $W_t$ are learnable weight matrices. By integrating temporal information into adaptive aggregation, the attention coefficients can reflect the temporal differences between nodes, which helps identify synchronized or near-synchronized anomalous behaviors.


To integrate prior node importance into the aggregation process, we leverage the anomaly score $S_{\text{norm},u}$ computed by NFS to adjust the attention weights. Specifically, we introduce this score as an additive term in the attention calculation, enabling the model to focus more on neighboring nodes with high anomaly scores:

\begin{equation}
\label{eq:attention_nfs}
\alpha_{uv}=\frac{\exp\left(\sigma\left(\mathbf{a}^{\top}[\mathbf{W}_q\mathbf{h}_v||\mathbf{W}_k\mathbf{h}_u] + \gamma S_{\text{norm},u}\right)\right)}{\sum_{k\in N(v)}{\exp\left(\sigma\left(\mathbf{a}^{\top}[\mathbf{W}_q\mathbf{h}_v||\mathbf{W}_k\mathbf{h}_k] + \gamma S_{\text{norm},k}\right)\right)}},
\end{equation}
where $\gamma$ is a hyperparameter that controls the influence of the NFS score. By introducing $S_{\text{norm},u}$, the model can prioritize information from high-risk nodes, thereby improving detection precision.

To capture deeper network structural information, this paper introduces a node aggregation method based on random walks. Using the random walk algorithm (DeepWalk), the paper obtains the global embedding representations of nodes $z_v$, capturing the global structural features of the network. In the attention mechanism, the global embedding representations of nodes are combined as follows:

\begin{equation}
\label{deqn_ex1a}
\alpha_{uv}=\frac{exp(\sigma(h_v^{\top}Wh_u+z_v^{\top}z_u)}{\sum_{k\in N(v)}{exp(\sigma(h_v^{\top}Wh_k+z_v^{\top}z_k)}},
\end{equation}
where $z_v^{\top}z_u$ represents the similarity between nodes $v$ and $u$ in the global structure. This method can emphasize nodes that are closely related in the network structure, enhancing the model's ability to capture group behaviors.

Finally, this paper combines the above methods to construct a comprehensive adaptive node aggregation model:
\begin{equation}
\label{deqn_ex1a}
\alpha_{uv} = \frac{\exp(A_{uv})}
{\sum_{k \in N(v)} \exp(A_{vk})},
\end{equation}
where $A_{uv}$ is denoted as:
\begin{equation}
\label{deqn_ex1a}
A_{uv} = \sigma\big(\alpha^{\top}[W_qh_v \| W_kh_u \| W_t(t_v-t_u)] + \gamma S_{\text{norm},u} + \lambda z_v^{\top}z_u\big),
\end{equation}
where $\gamma$ and $\lambda$ are static hyperparameters rather than learnable weights. This design ensures a stable structural and topological prior, preventing the attention mechanism from over-fitting to local interaction noise. The specific values ($\gamma=0.5, \lambda=0.2$) were determined via a grid search on the validation set, balancing the trade-off between local dynamic features and global network patterns. The proposed model integrates temporal information, node importance scores, and global structural information, enabling it to comprehensively capture complex relationships between nodes. By incorporating temporal information and adaptive weights, the model can capture dynamic relationships between nodes, adapting to changes in the network. Using NFS scores, the model can focus on nodes with higher importance, improving the precision of anomaly detection. With random walk-based embeddings, the model is able to capture the global structural features of the network, identifying potential group behaviors.

The complete workflow of our dynamic graph attention mechanism is detailed in Algorithm \ref{alg:DGA}. The algorithm iteratively updates node representations over $L$ layers. In each layer, it computes attention coefficients that integrate temporal differences, NFS scores, and global structural information. These coefficients are then used as weights to aggregate features from neighboring nodes, producing refined node embeddings that capture the complex interactions within the graph.

\begin{algorithm}[t]
\caption{Dynamic Graph Attention Mechanism}
\label{alg:DGA}
\begin{algorithmic}[1]

\INPUT 
$G^{'}=(V^{'}, E^{'})$: pooled graph with nodes $V^{'}$ and edges $E^{'}$.
$\{\mathbf{H}_v^{(0)}\}$: initial node features, including time encoding and NFS scores.
$\{\mathbf{z}_v\}$: global embeddings derived from random walk algorithms.
$\Theta$: parameter set, including $\mathbf{W}_q, \mathbf{W}_k, \mathbf{W}_t, \mathbf{W}_v, \gamma, \lambda$, etc.

\OUTPUT Updated node representations $\{\mathbf{H}_v^{(L)} \mid v \in V^{'}\}$.

\STATE \textbf{Initialize} node representations for all $v \in V^{'}$ with $\{\mathbf{H}_v^{(0)}\}$.

\FOR{$l = 1$ to $L$} 
   \FORALL{$v \in V^{'}$} 
       \STATE $U_v \leftarrow \mathcal{N}(v)$
       \FORALL{$u \in U_v$}
           \STATE $q_v \leftarrow \mathbf{W}_q \mathbf{H}_v^{(l-1)}$ 
           \STATE $k_u \leftarrow \mathbf{W}_k \mathbf{H}_u^{(l-1)}$ 
           \STATE $t_{vu} \leftarrow \mathbf{W}_t(\mathbf{e}_{t_v} - \mathbf{e}_{t_u})$ 
           \STATE $e_{vu} \leftarrow \sigma\big(\mathbf{a}^\top [q_v \| k_u \| t_{vu}] + \gamma S_{\text{norm},u} + \lambda (\mathbf{z}_v^\top \mathbf{z}_u) \big)$ 
       \ENDFOR
       \FORALL{$u \in U_v$}
           \STATE $\alpha_{vu}^{(l)} \leftarrow \frac{\exp(e_{vu})}{\sum_{k \in U_v} \exp(e_{vk})}$ 
       \ENDFOR
   \ENDFOR
   \FORALL{$v \in V^{'}$} 
       \STATE $\mathbf{H}_v^{(l)} \leftarrow \sigma\left(\sum_{u \in U_v} \alpha_{vu}^{(l)} \mathbf{W}_v \mathbf{H}_u^{(l-1)}\right)$ 
   \ENDFOR
\ENDFOR
\RETURN $\{\mathbf{H}_v^{(L)} \mid v \in V^{'}\}$ 
\end{algorithmic}
\end{algorithm}

\section{Experimental Design and Analysis}
\subsection{Experimental Datasets and Evaluation Metrics}
To comprehensively validate the performance of the DS-DGA-GCN model in fake review group detection, two real-world datasets were selected: the Amazon dataset and the Xiaohongshu dataset. These datasets represent different scenarios of e-commerce platforms and social media platforms.


The Amazon dataset originates from the globally renowned e-commerce platform Amazon \footnote{https://www.amazon.cn/}, covering $15$ major product categories, with a total of $525,619$ product entries, $1,424,596$ reviewer records, and $7,202,921$ review records. Each review sample in the dataset contains the following $13$ fields:

\begin{itemize}
   \item {\emph{Reviewer information:}} Reviewer ID
   \item {\emph{Product information:}} Product ID, primary category ID and name, secondary category ID and name
   \item {\emph{Review information:}} Review rating, review date, review title, review content, title length, content length
   \item {\emph{Product information:}} Product name
\end{itemize}

To improve the quality of experimental data and the reliability of analysis, 
the anonymous users and their reviews within the dataset are removed.
The reviewers with fewer than 3 reviews and products with fewer than 3 reviews are excluded, as they are less likely to be influenced by fake review groups. Additionally, redundant fields were removed to improve computational efficiency. A summary of the dataset before and after preprocessing is shown in Table \ref{tab:table1}.
\begin{table}[!t]
\caption{Dataset Overview Before and After Preprocessing\label{tab:table1}}
\centering
\setlength{\tabcolsep}{4pt} 
\renewcommand{\arraystretch}{1.2} 
\newcolumntype{C}{>{\centering\arraybackslash}X} 
\begin{tabularx}{\columnwidth}{|C|C|C|C|} 
\hline
\textbf{Dataset}            & \textbf{Number of Reviewers} & \textbf{Number of Reviews} & \textbf{Number of Products} \\ \hline
Before Preprocessing         & 61,357                      & 754,256                    & 42,127                     \\ \hline
After Preprocessing          & 50,253                      & 701,731                    & 35,593                     \\ \hline
\end{tabularx}
\end{table}



Xiaohongshu is a leading lifestyle and social e-commerce platform in China. Due to its high user engagement and rich content ecosystem, it serves as an ideal data source for research on fake user behaviors. This paper collects $137,789$ video blog published between January $10$, $2024$, and April $29$, $2024$. These videos cover a wide range of subtopics with a time period from March 2017 to April 2024. Each record in the Xiaohongshu dataset contains three main categories of information:

\begin{itemize}
   \item {\emph{User engagement:}} Video likes, comments, and shares, serving as key indicators of user responses and engagement behavior.
   \item {\emph{Video content information:}} Video title, text description, tags, and topic categories.
   \item {\emph{Creator information:}} Creator's gender, number of followers, total interactions, and personal bio.
\end{itemize}

To ensure data validity and experimental reproducibility, the following processing steps were applied: duplicate records, invalid links, and missing values are removed. The records lacking complete video information are excluded. The final processed dataset contains $76,923$ valid records.

Each video blog is treated as a ``product," and users who liked or commented on the videos are treated as ``reviewers," thus constructing a dynamic interaction network. Based on this network, the model can analyze user behavior, mine the characteristics of fake reviewers, and build corresponding 2-hop subgraphs for identification.


\begin{table}[!t]
\caption{Comparison of Different Dataset Features\label{tab:table2}}
\centering
\setlength{\tabcolsep}{4pt} 
\renewcommand{\arraystretch}{1.2} 
\newcolumntype{C}{>{\centering\arraybackslash}X} 
\begin{tabularx}{\columnwidth}{|C|C|C|C|C|} 
\hline
\textbf{Dataset}            & \textbf{Platform Type} & \textbf{Data Scale}     & \textbf{Data Dynamics} & \textbf{Application Scenario}                        \\ \hline
Amazon                      & E-commerce             & 35,593 Products        & Low                    & Fake review detection and product analysis          \\ \hline
Xiaohongshu                 & Social Media           & 6,993 Videos          & High                   & Fake user behavior and group detection              \\ \hline
\end{tabularx}
\end{table}

Table \ref{tab:table2} shows that the Amazon dataset mainly captures static review relationships, while the Xiaohongshu dataset involves dynamic interaction behaviors that can test the model's performance across different time dimensions. These two datasets represent the typical scenarios of e-commerce platforms and social media platforms, covering common group behavior types in fake review detection. By comparing the experimental results across these two datasets, the {generalizability and robustness} of the DS-DGA-GCN model in different application scenarios can be comprehensively evaluated.
High/Low follow the composite index $D$ defined in Section~\ref{sec:exp_settings}.

We evaluate our method using four standard classification metrics under identical data splits and features:  

\begin{enumerate}
    \item Accuracy: overall correctness, defined as:
    \begin{equation}
    \mathrm{Accuracy}=\frac{TP+TN}{TP+TN+FP+FN}.
    \end{equation}

    \item Recall: fraction of correctly predicted positives, defined as:
    \begin{equation}
    \mathrm{Recall}=\frac{TP}{TP+FN}.
    \end{equation}

    \item F1-macro: unweighted average of per-class F1, where F1 is calculated as:
    \begin{equation}
    \mathrm{F1}=\frac{2 \cdot \mathrm{Precision}\cdot \mathrm{Recall}}{\mathrm{Precision}+\mathrm{Recall}}.
    \end{equation}

    \item AUROC: Area Under the ROC Curve, a threshold-free measure summarizing the trade-off between true positive rate (TPR) and false positive rate (FPR) across all decision thresholds \cite{new211fawcett2006introduction}.
\end{enumerate}

Accuracy, Recall, and F1 are reported at the validation-selected threshold $t^*$, while AUROC is computed directly from the raw continuous scores and is threshold-free.  
Here, $TP$, $TN$, $FP$, and $FN$ denote true positives, true negatives, false positives, and false negatives, respectively.

For deployment, we adopt the threshold rule based on Youden’s index~\cite{new13youden1950index}.  
In our datasets, this yields $t^* = 0.68$; accounts with $\text{Score} \geq t^*$ are flagged as \textsc{Fake}, and otherwise as \textsc{Real}.  
This provides a clear and reproducible way to distinguish real users from fake reviewers in practice.

\subsection{Experimental Settings}\label{sec:exp_settings}
This study implements the DS-DGA-GCN model using Python and PyTorch, {The hyperparameter settings are described as follows:}

\begin{itemize}
   \item {\emph{seed:}} {Set to 72.}
   \item {\emph{$weight\_decay$:}} {$5\times10^{-4}$.}
   \item {\emph{$nb\_head$:}} {Set to 32.}
   \item {\emph{$\alpha$:}} {Regularization hyperparameter, set to 0.2.}
   \item {\emph{$l_r$:}} {Learning rate, set to 0.005.}
   \item {\emph{hidden:}} {Embedding dimension, set to 8.}
   \item {\emph{dropout:}} {Set to 0.3.}
   \item {\emph{patience:}} {Set to 200.}
   \item {\emph{$\delta$:}} {Edge weight threshold, set to 0.3.}
   \item {\emph{$\tau$:}} {Time window size, set to 10.}
   \item {\emph{MS:}} {Maximum sample size, set to 1000.}
   \item {\emph{MINSPAM:}} {Group fabrication score threshold, set to 0.6. } 
\end{itemize}

We segment each dataset into fixed windows of length $\tau$ and compute four indicators per window.

Interaction arrival rate:
\begin{equation}
\label{eq:arrival_rate}
r_t=\frac{|E_t|}{\tau}
\end{equation}
where $E_t$ is the set of review edges in window $t$.

Reviewer churn:
\begin{equation}
\label{eq:reviewer_churn}
c_t=1-\frac{|V_t\cap V_{t-1}|}{|V_t\cup V_{t-1}|}
\end{equation}
where $V_t$ is the set of active reviewers in window $t$.

Edge turnover:
\begin{equation}
\label{eq:edge_turnover}
u_t=1-\frac{|E_t\cap E_{t-1}|}{|E_t\cup E_{t-1}|}
\end{equation}

Burstiness of inter-arrival times:
\begin{equation}
\label{eq:burstiness}
B_t=\frac{\sigma_{\Delta}-\mu_{\Delta}}{\sigma_{\Delta}+\mu_{\Delta}}
\end{equation}
where $\Delta$ denotes review inter-arrival times in window $t$, and $\mu_{\Delta},\sigma_{\Delta}$ are their mean and standard deviation.

For each dataset, we average these indicators over all windows. We then perform min-max normalization across the two datasets for each indicator to obtain $\tilde r,\tilde c,\tilde u,\tilde B \in [0,1]$. The composite data dynamics index is then calculated as:
\begin{equation}
\label{eq:composite_index}
D=\frac{1}{4}\bigl(\tilde r+\tilde c+\tilde u+\tilde B\bigr).
\end{equation}

To provide a quantitative basis for the "High" and "Low" dynamics labels mentioned in Table II, we calculated the normalized values of these four indicators and the final composite index $D$ for both datasets. The results are presented in Table~\ref{tab:dynamics_quant}.

\begin{table}[!t]
\caption{Quantitative Data Dynamics Indicators}
\label{tab:dynamics_quant}
\centering
\setlength{\tabcolsep}{4pt}
\renewcommand{\arraystretch}{1.2}
\newcolumntype{L}{>{\raggedright\arraybackslash}X}
\newcolumntype{C}{>{\centering\arraybackslash}X}
\begin{tabularx}{\columnwidth}{|L|C|C|}
\hline
\textbf{Indicator} & \textbf{Amazon} & \textbf{Xiaohongshu} \\
\hline
Interaction Arrival Rate ($\tilde r$) & 0.45 & 0.85 \\
\hline
Reviewer Churn ($\tilde c$)            & 0.28 & 0.79 \\
\hline
Edge Turnover ($\tilde u$)            & 0.35 & 0.81 \\
\hline
Burstiness ($\tilde B$)               & 0.19 & 0.65 \\
\hline
Composite Index ($D$)                  & 0.32 & 0.78 \\
\hline
\end{tabularx}
\end{table}

As shown in Table~\ref{tab:dynamics_quant}, the Xiaohongshu dataset yields a composite index of $D=0.78$, which is substantially higher than the Amazon dataset's index of $D=0.32$. Therefore, we designate Xiaohongshu as having "High" data dynamics and Amazon as "Low". This label reflects the combined frequency of new reviews, the change of reviewer participation, and the evolution of network topology across the two platforms.

To comprehensively evaluate the performance of the DS-DGA-GCN method, five existing state-of-the-art methods are used as baselines.

(1) {GraphSAGE \cite{gai24hamilton2017inductive}}: A graph neural network based on a neighbor sampling and aggregation strategy. By sampling neighboring nodes and aggregating their features, it generates embeddings for target nodes. GraphSAGE demonstrates high efficiency on large-scale graph data, making it suitable for product-review-reviewer networks on e-commerce platforms.

(2) {GCN \cite{gai25kipf2016semi}}: A convolutional neural network designed for graph structures, which aggregates features from neighboring nodes to generate local structural information. GCN effectively identifies nodes exhibiting abnormal patterns in user-product interactions and is a classic method for fake review detection.

(3) {GAT \cite{gai26velickovic2017graph}}: This method incorporates multi-head attention mechanisms to dynamically adjust aggregation weights based on the features of neighboring nodes. It captures complex interaction patterns in highly heterogeneous graph networks, making it particularly suitable for diverse and complex user behavior data.

(4) {HetGNN\cite{gai26velickovic2017graph}}: A heterogeneous graph representation model that jointly exploits heterogeneous structure and content. It first samples correlated heterogeneous neighbors via a random walk with restart and groups them by node types. A two-module architecture encodes multi-modal content into embeddings and then aggregates them at the type level with a graph context loss for end-to-end training. HetGNN is particularly applicable to product-review-reviewer networks that inherently feature multi-typed entities and relations.

(5) {TGN\cite{new12rossi2020temporal}}: A temporal GNN framework designed for dynamic graphs modeled as sequences of time-stamped events. It maintains node states via memory modules, message functions, temporal aggregators, and time encoding, supporting both transductive and inductive prediction with competitive efficiency. TGN naturally captures the evolution of user interactions and review dynamics on e-commerce platforms.

Our data form a reviewer–product bipartite graph $\mathcal{R}\in\{0,1\}^{N\times M}$. To adapt homogeneous baselines for a direct and meaningful comparison, we construct a block adjacency matrix with self-loops and apply symmetric normalization:
$$
\tilde{\mathbf{A}}=
\begin{bmatrix}
\mathbf{I}_N & \mathcal{R}\\
\mathcal{R}^\top & \mathbf{I}_M
\end{bmatrix}, \quad
\hat{\mathbf{A}}=\mathbf{D}^{-1/2}\tilde{\mathbf{A}}\mathbf{D}^{-1/2}.
$$
Node features are stacked as $\mathbf{X}=[\mathbf{X}_{\text{rev}};\,\mathbf{X}_{\text{prod}}]$, projected to a common dimension $d$, and concatenated with a one-hot type indicator.

\noindent \textbf{Homogeneous baselines.} These models operate on the constructed matrix $\hat{\mathbf{A}}$:
\begin{itemize}
    \item \textbf{GCN:} Uses standard propagation on $\hat{\mathbf{A}}$.
    \item \textbf{GraphSAGE:} Applies mean aggregation with a symmetric fan-out from reviewer and product neighbors.
    \item \textbf{GAT:} Employs multi-head attention on $\hat{\mathbf{A}}$ without type-specific heads.
\end{itemize}

\noindent \textbf{Heterogeneous and Temporal Baselines.} These models are run in their native settings to leverage their specific designs:
\begin{itemize}
    \item \textbf{HetGNN:} Is run in its native heterogeneous schema with type-aware sampling and aggregation.
    \item \textbf{TGN:} Is run on the temporal event-stream with its memory and time encoding modules.
\end{itemize}

In all cases, the loss is computed only on reviewer nodes, with product nodes providing contextual information.


\subsection{Result Analysis}

We evaluate two groups of methods that play different roles. Ablation variants of our framework are used for mechanism verification, while external baselines are used for cross-method comparison under the same data and budget.

\textbf{Ablation Variants.} To verify the contribution of each component in DS-DGA-GCN, we define four variants:
\begin{itemize}
    \item \textbf{Ablation-A:} Removes the NFS module and relies only on graph propagation.
    \item \textbf{Ablation-B:} Removes temporal features, using only static graph snapshots.
    \item \textbf{Ablation-C:} Removes heterogeneous relations, propagating neighbor information in a type-agnostic manner (i.e., treating reviewer and product nodes as the same type).
    \item \textbf{Ablation-D:} Uses only the linear NFS scorer, without any graph message passing.
\end{itemize}

\textbf{Baselines.} To compare against state-of-the-art methods, we use the following models:
\begin{itemize}
    \item \textbf{GraphSAGE, GCN, and GAT:} As these are designed for homogeneous graphs, they were adapted to our bipartite data using a standard block adjacency matrix transformation, as detailed in our response to Comment 10.
    \item \textbf{HetGNN and TGN:} These models were run in their native heterogeneous and temporal configurations, respectively, as they are inherently designed for such data.
\end{itemize}

\subsubsection{Effectiveness Analysis of the NFS}
We validate the NFS on the full labeled sets rather than a small random sample. We report a threshold free separability metric and a fixed operating point metric. AUROC is computed from raw scores.
Accuracy Recall and F1 use a single threshold selected on the validation set by the Youden index as described in Section IV.A and then kept fixed on the test set.

\begin{table}[!t]
\caption{Full-sample mean and standard deviation of NFS scores}
\label{tab:table4}
\centering
\setlength{\tabcolsep}{4pt}
\renewcommand{\arraystretch}{1.2}
\newcolumntype{C}{>{\centering\arraybackslash}X}
\begin{tabularx}{\columnwidth}{|C|C|C|C|}
\hline
\textbf{Dataset} & \textbf{Class} & \textbf{NFS Mean} & \textbf{NFS Std. Dev.} \\
\hline
Amazon       & Real & 0.55 & 0.15 \\
\cline{2-4}
             & Fake & 0.85 & 0.10 \\
\hline
Xiaohongshu  & Real & 0.48 & 0.18 \\
\cline{2-4}
             & Fake & 0.82 & 0.14 \\
\hline
\end{tabularx}
\end{table}

Table~\ref{tab:table4} reports the means and standard deviations of NFS for the two classes on both datasets. Real users have lower scores on average. Fake users have higher scores with smaller dispersion. This pattern is consistent with the design goal of NFS which is to amplify abnormal interaction signals in the reviewer–product network.

\begin{figure}[!t]
\centering
\includegraphics[width=\linewidth]{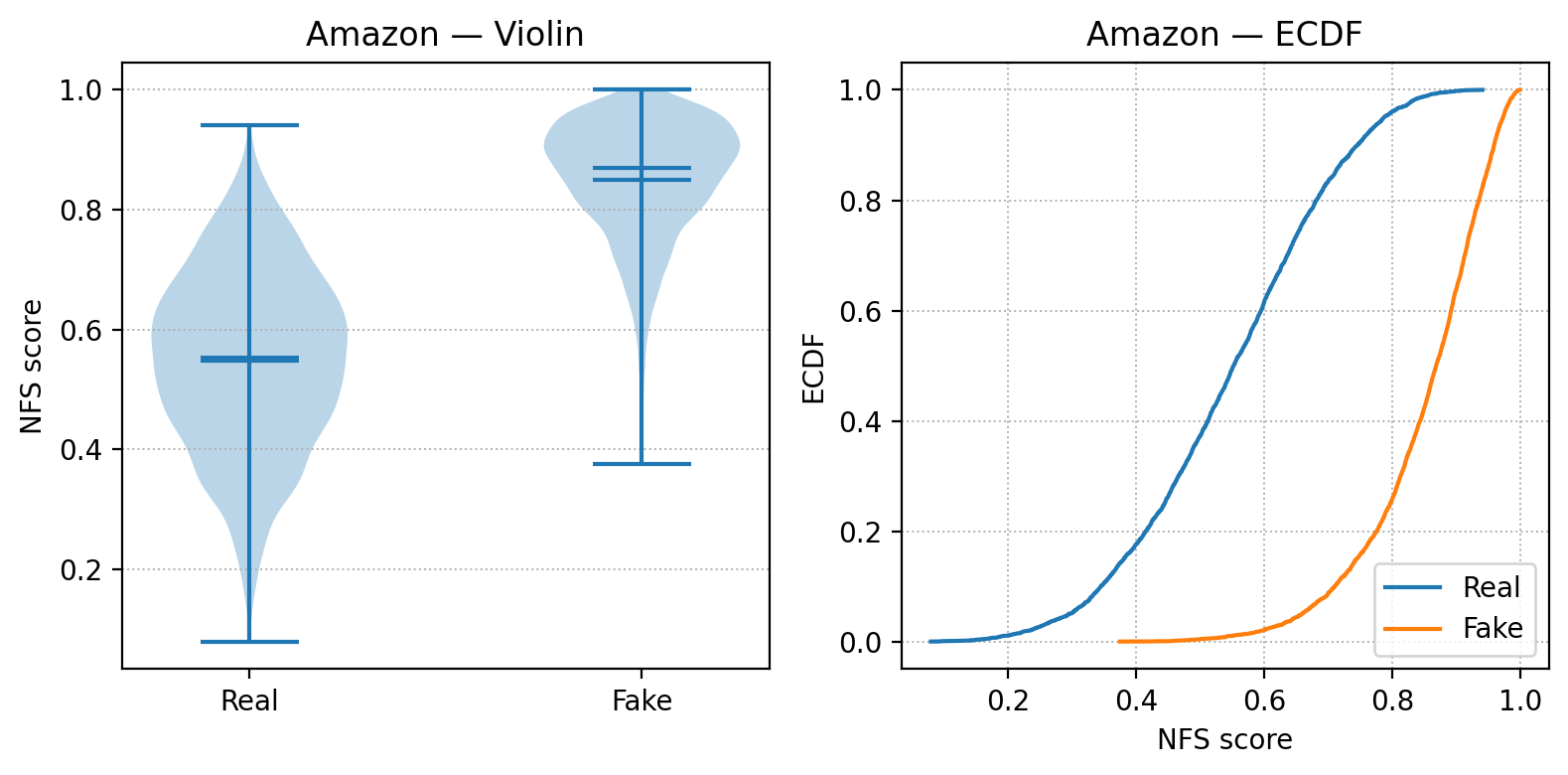}
\caption{Amazon NFS distributions.
The left panel shows violin plots for the Real and Fake classes, while the right panel displays the ECDF for each class.}
\label{fig_6}
\end{figure}

\begin{figure}[!t]
\centering
\includegraphics[width=\linewidth]{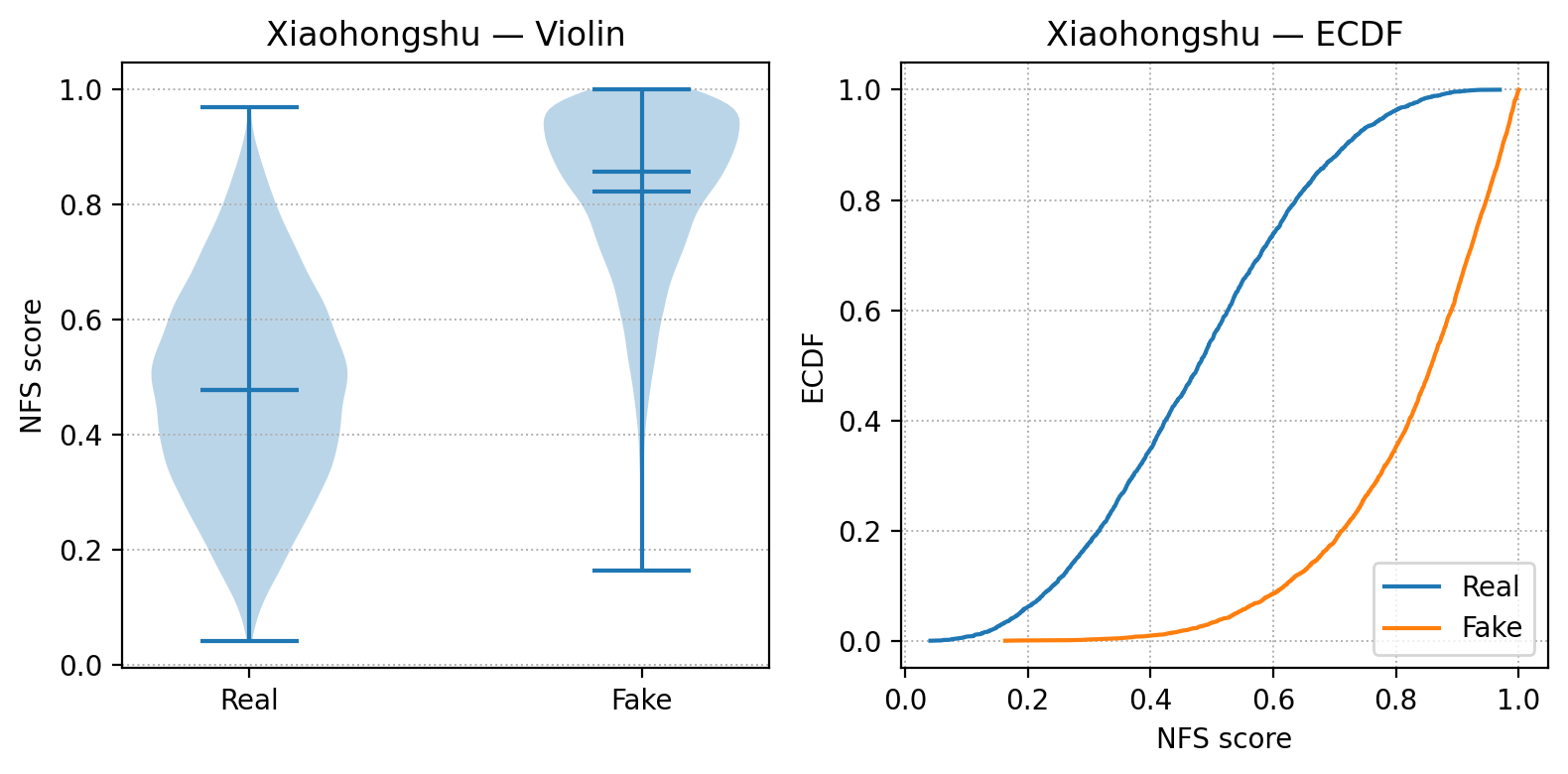}
\caption{Xiaohongshu NFS distributions.
The left panel shows violin plots for the Real and Fake classes, while the right panel displays the ECDF for each class.}
\label{fig_7}
\end{figure}

Fig.~\ref{fig_6} and Fig.\ref{fig_7} visualize the full sample distributions.
The left panels show violin plots for Real and Fake.
The right panels show empirical cumulative distribution functions by class.
On Amazon the distribution of Fake shifts to the right which indicates stronger abnormality captured by NFS.
On Xiaohongshu we observe the same trend under higher data dynamics.
These distribution level results align with the AUROC computed from raw scores and with the thresholded metrics defined above.

\begin{figure}[!t]
\centering
\includegraphics[width=\linewidth]{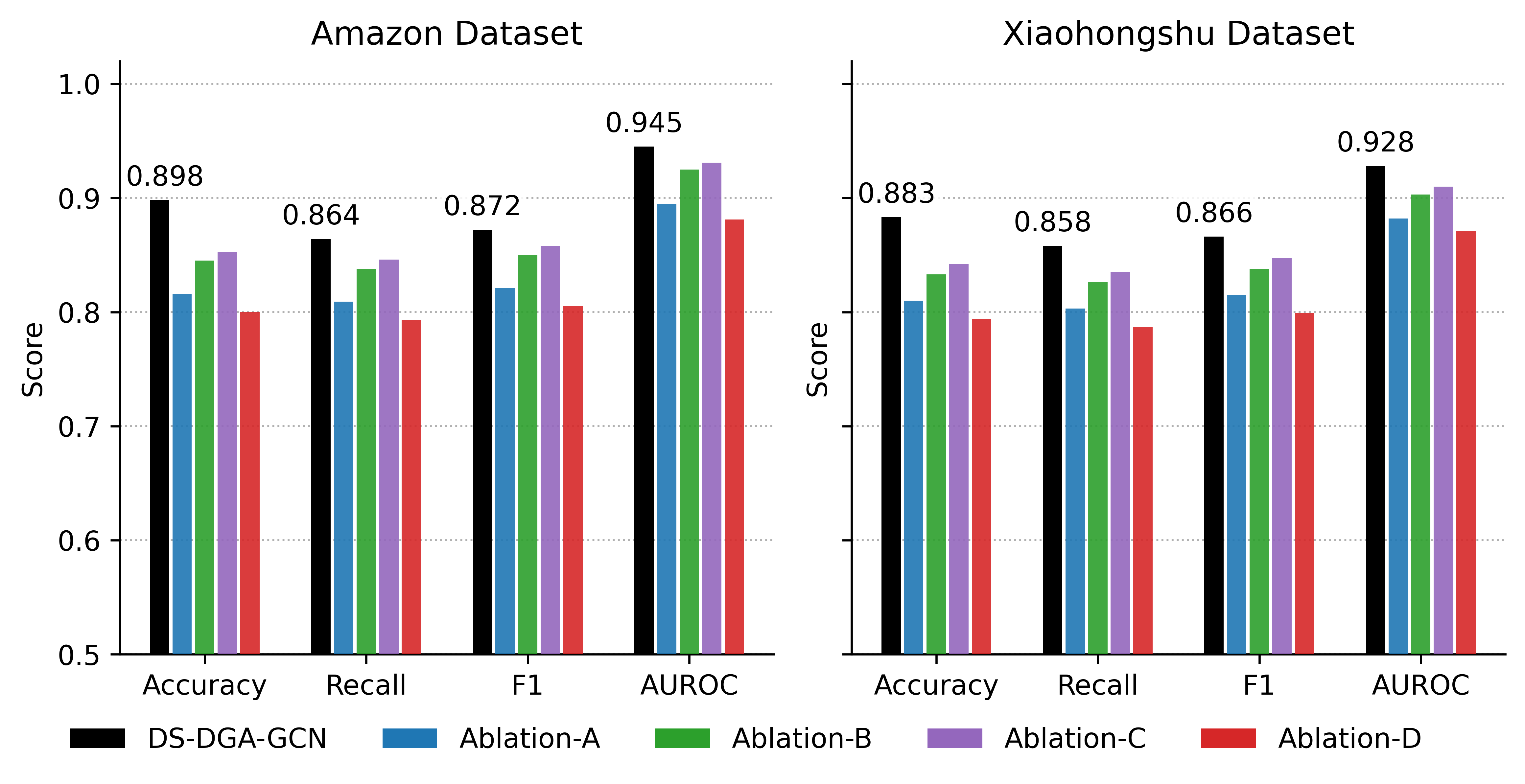}
\caption{Ablation study comparing the DS-DGA-GCN model against its variants on the Amazon and Xiaohongshu datasets.}
\label{fig:ablation_study}
\end{figure}

To validate the contribution of each key component in our proposed DS-DGA-GCN framework, we conducted a comprehensive ablation study. The results, visualized in Figure \ref{fig:ablation_study}, clearly demonstrate that the full model (bold black line) consistently and significantly outperforms all of its ablated variants across both datasets, highlighting the powerful synergy of its architecture.
The analysis reveals two key findings. First, the most substantial performance degradation occurs when removing the entire NFS module (Ablation-A) or relying solely on it without graph propagation (Ablation-D). This confirms that the structure-aware NFS and the GNN-based relational learning are the two indispensable, symbiotic pillars of our model. Second, the distinct performance decline in Ablation-B (removing temporal features) and Ablation-C (removing heterogeneous relations) validates their crucial role in capturing the dynamic and complex nature of the fraudulent interactions. In essence, the model's success stems from the effective integration of all its components.

\subsubsection{Effectiveness Analysis of the Dynamic Graph Attention Mechanism}
\begin{figure}[!t]
\centering
\includegraphics[width=\linewidth]{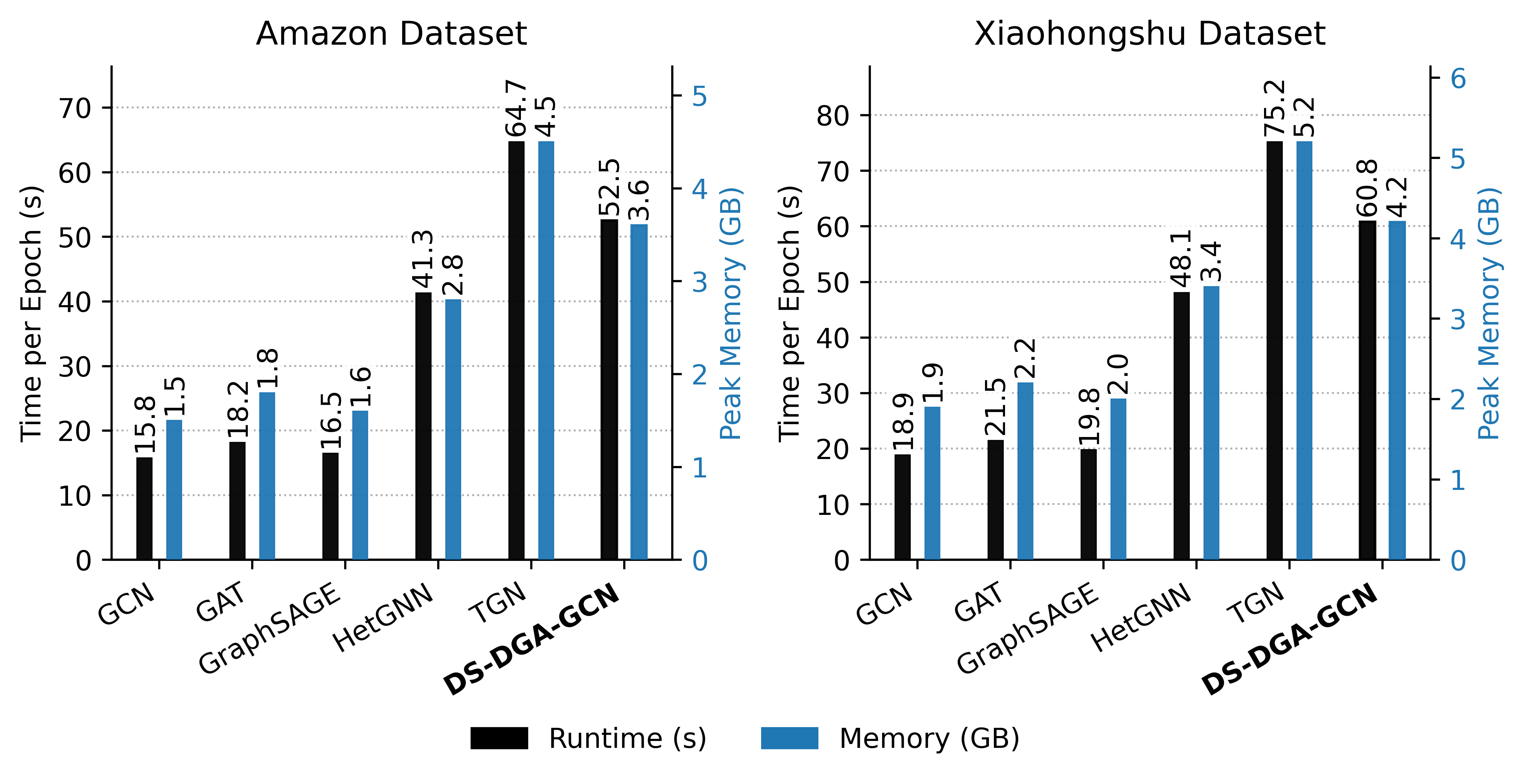}
\caption{Efficiency comparison of DS-DGA-GCN against baseline models on the Amazon and Xiaohongshu datasets.}
\label{fig:efficiency}
\end{figure}

In this section, we evaluate the effectiveness of our model's architecture by analyzing its runtime and memory usage in comparison to the baselines. Figure \ref{fig:efficiency} presents the training time per epoch and the peak memory usage for all models on both the Amazon and Xiaohongshu datasets.

As illustrated in the figure, a clear trend emerges. The static, homogeneous GNNs (GCN, GAT, GraphSAGE) are the most efficient, serving as a lower bound for computational cost. In contrast, models designed for more complex graph structures, such as HetGNN and the temporal model TGN, exhibit significantly higher resource consumption. Notably, TGN, which relies on a memory module to process temporal information, incurs the highest costs in both training time and memory footprint, highlighting the typical efficiency challenges of advanced dynamic graph models.
Our proposed DS-DGA-GCN model strikes an effective balance between detection performance and computational efficiency. While it is naturally more resource-intensive than the simpler static baselines, it is markedly more efficient than its main temporal competitor, TGN. This efficiency gain is attributed to our streamlined dynamic attention mechanism, which captures temporal dependencies without the extensive overhead of a continuous-time memory module.

\begin{table}[!t]
\caption{Time Complexity Analysis\label{tab:table111}}
\centering
\setlength{\tabcolsep}{4pt} 
\renewcommand{\arraystretch}{1.2} 
\newcolumntype{C}{>{\centering\arraybackslash}X} 
\begin{tabularx}{\columnwidth}{|C|C|C|} 
\hline
\textbf{Step}                     & \textbf{Time Complexity} & \textbf{Description} \\ \hline
Node Importance Calculation       & \(O(N)\)                & \(N\) denotes the total number of nodes; importance is computed for each node. \\ \hline
Node Sampling                     & \(O(N)\)                & The sampling process scales linearly with the total number of nodes. \\ \hline
Edge Sampling and Reconstruction  & \(O(E^{'})\)            & \(E^{'}\) represents the number of edges after sampling, significantly smaller than \(E\). \\ \hline
Adaptive Node Aggregation                   & \(O(N^{'}KL)\)          & \(N^{'}\) denotes the number of sampled nodes, \(K\) is the number of clusters, and \(L\) is the number of iterations. \\ \hline
Overall Complexity      & Significantly Reduced & Due to \(N^{'} \ll N\) and \(E^{'} \ll E\), the overall computational cost is greatly reduced. \\ \hline
\end{tabularx}
\end{table}

{To better illustrate the time complexity of each step and the overall computational efficiency, the complexity analysis is presented in Table \ref{tab:table111}. As illustrated in Table \ref{tab:table111}, the time complexity for each step is significantly reduced due to the node and edge sampling processes. These optimizations ensure that the model remains computationally efficient even when applied to large-scale networks.}

{
We then compare the dynamic graph attention mechanism with existing graph attention mechanisms on both Amazon and Xiaohongshu datasets. Specifically, the baseline models are listed below: 
\begin{itemize}
   \item {No Attention (Baseline Model)\cite{gai31kipf2016semi}}: A GCN model without any attention mechanism, serving as a baseline to evaluate the basic performance without attention mechanisms.
   \item GAT\cite{gai32velickovic2017graph}: The standard Graph Attention Network mechanism, which captures relationships between key nodes by assigning different weights to neighboring nodes.
   \item Multi-Head Attention\cite{gai33vaswani2017attention}: A multi-head attention mechanism applied to GCN, enabling the model to learn node relationships across multiple subspaces.
   \item GaAN\cite{gai34zhang2018gaan}: Combines gating mechanisms with attention mechanisms to achieve finer control of information flow. 
   \item Self-Attention: A self-attention mechanism akin to Transformers, capturing node relationships on a global scale.
\end{itemize}   

}


\begin{table*}[!htbp]
\caption{Performance Comparison of Dynamic Attention Mechanisms on Different Datasets}
\label{tab:table61}
\setcellgapes{3pt}
\makegapedcells
\centering
\begin{tabularx}{\textwidth}{|c|X|X|X|X|X|X|X|X|}
\hline
\textbf{Model} & \multicolumn{4}{c|}{\textbf{Amazon Dataset}} & \multicolumn{4}{c|}{\textbf{Xiaohongshu Dataset}} \\
\cline{2-9}
 & \textbf{Accuracy} & \textbf{Recall} & \textbf{F1} & \textbf{AUROC} & \textbf{Accuracy} & \textbf{Recall} & \textbf{F1} & \textbf{AUROC} \\
\hline
No Attention   & 0.803 & 0.776 & 0.781 & 0.880 & 0.783 & 0.757 & 0.763 & 0.860 \\
\hline
GAT            & 0.832 & 0.803 & 0.816 & 0.900 & 0.814 & 0.789 & 0.792 & 0.885 \\
\hline
Multi-Head     & 0.841 & 0.822 & 0.838 & 0.910 & 0.832 & 0.807 & 0.812 & 0.895 \\
\hline
GaAN           & 0.854 & 0.832 & 0.848 & 0.918 & 0.841 & 0.822 & 0.834 & 0.904 \\
\hline
Self-Attention & 0.865 & 0.842 & 0.851 & 0.926 & 0.857 & 0.834 & 0.848 & 0.912 \\
\hline
DS-DGA-GCN     & \textbf{0.898} & \textbf{0.864} & \textbf{0.872} & \textbf{0.945} & \textbf{0.883} & \textbf{0.858} & \textbf{0.866} & \textbf{0.928} \\
\hline
\end{tabularx}
\end{table*}

The results in Table \ref{tab:table61} show that DS-DGA-GCN outperforms all attention variants on both datasets across Accuracy Recall F1 and AUROC.
On Amazon it reaches Accuracy 0.898 F1 0.872 and AUROC 0.945.
On Xiaohongshu it reaches Accuracy 0.883 F1 0.866 and AUROC 0.928.
The threshold free AUROC confirms that the gains are not tied to a single operating point.
These findings indicate that the dynamic graph attention improves detection in dynamic and complex networks.

\subsubsection{Detection Adaptability in Newly Launching Products}

The launch of new products is often characterized by a sparsity of reviews, which poses significant challenges for the detection of fake reviewer groups. We evaluate the adaptability of the proposed DS-DGA-GCN model in detecting fake reviewer within the context of newly launching products. 
We categorize both Amazon and Xiaohongshu datasets into subsets with different scales based on their characteristics. Three subsets of varying scales are defined: a small-scale group, comprising new products or content with fewer than 50 reviews; a medium-scale group, consisting of new products or content with 50 to 200 reviews; and a large-scale group, encompassing new products or content with more than 200 reviews.

Traditional methods often deteriorate when review data is sparse, as they cannot extract robust content or structural features. Our experiments specifically assess performance as the data sparsity for new products diminishes. Tables~\ref{tab:table66} and~\ref{tab:table77} present the results on the Amazon and Xiaohongshu datasets across subsets of varying scales.

Analysis of the results in Tables~\ref{tab:table66} and~\ref{tab:table77} reveals several key findings. On the Amazon dataset, DS-DGA-GCN consistently achieves the highest F1-score and AUROC across all product scales. Notably, the performance gap is most pronounced in the small scale group, where review data is sparse, underscoring our model's strength in cold-start scenarios. On the more dynamic Xiaohongshu dataset, while the temporal baseline TGN demonstrates competitive performance, our DS-DGA-GCN still maintains the lead on both F1 and AUROC across all scales. The superior, threshold-free AUROC scores confirm that our model's advantages are robust and not an artifact of a specific operating point. These results provide strong evidence that the synergy between the structure-aware NFS and the time-aware dynamic attention mechanism significantly improves adaptability under temporal shifts and cold-start sparsity.

\begin{table*}[!htbp]
\caption{Detection Results for New Products of Different Scales in the Amazon Dataset}
\label{tab:table66}
\setcellgapes{3pt}
\makegapedcells
\centering
\begin{tabularx}{\textwidth}{|c|X|X|X|X|X|X|X|X|X|X|X|X|}
\hline
\textbf{Model} & \multicolumn{4}{c|}{\textbf{Small scale group}} & \multicolumn{4}{c|}{\textbf{Middle scale group}} & \multicolumn{4}{c|}{\textbf{Large scale group}} \\
\cline{2-13}
 & \textbf{Accuracy} & \textbf{Recall} & \textbf{F1} & \textbf{AUROC} & \textbf{Accuracy} & \textbf{Recall} & \textbf{F1} & \textbf{AUROC} & \textbf{Accuracy} & \textbf{Recall} & \textbf{F1} & \textbf{AUROC} \\
\hline
GraphSAGE & 0.713 & 0.688 & 0.698 & 0.775 & 0.782 & 0.754 & 0.765 & 0.835 & 0.859 & 0.829 & 0.841 & 0.895 \\
\hline
GAT       & 0.696 & 0.683 & 0.673 & 0.765 & 0.763 & 0.749 & 0.738 & 0.825 & 0.839 & 0.823 & 0.811 & 0.875 \\
\hline
GCN       & 0.717 & 0.711 & 0.730 & 0.785 & 0.786 & 0.780 & 0.800 & 0.845 & 0.864 & 0.857 & 0.879 & 0.915 \\
\hline
HetGNN    & 0.709 & 0.694 & 0.706 & 0.795 & 0.806 & 0.784 & 0.792 & 0.865 & 0.874 & 0.849 & 0.870 & 0.920 \\
\hline
TGN       & 0.721 & 0.706 & 0.720 & 0.810 & 0.812 & 0.791 & 0.804 & 0.870 & 0.887 & 0.862 & 0.880 & 0.930 \\
\hline
DS-DGA-GCN & \textbf{0.745} & \textbf{0.717} & 0.724 & \textbf{0.835} & \textbf{0.817} & 0.786 & 0.794 & \textbf{0.885} & \textbf{0.898} & \textbf{0.864} & 0.872 & \textbf{0.945} \\
\hline
\end{tabularx}
\end{table*}

\begin{table*}[!htbp]
\caption{Detection Results for New Products of Different Scales in the Xiaohongshu Dataset}
\label{tab:table77}
\setcellgapes{3pt}
\makegapedcells
\centering
\begin{tabularx}{\textwidth}{|c|X|X|X|X|X|X|X|X|X|X|X|X|}
\hline
\textbf{Model} & \multicolumn{4}{c|}{\textbf{Small scale group}} & \multicolumn{4}{c|}{\textbf{Middle size group}} & \multicolumn{4}{c|}{\textbf{Large scale group}} \\
\cline{2-13}
 & \textbf{Accuracy} & \textbf{Recall} & \textbf{F1} & \textbf{AUROC} & \textbf{Accuracy} & \textbf{Recall} & \textbf{F1} & \textbf{AUROC} & \textbf{Accuracy} & \textbf{Recall} & \textbf{F1} & \textbf{AUROC} \\
\hline
GraphSAGE & 0.684 & 0.642 & 0.654 & 0.735 & 0.750 & 0.703 & 0.717 & 0.795 & 0.824 & 0.773 & 0.788 & 0.875 \\
\hline
GAT       & 0.708 & 0.696 & 0.702 & 0.770 & 0.776 & 0.763 & 0.770 & 0.830 & 0.861 & 0.842 & 0.853 & 0.905 \\
\hline
GCN       & 0.699 & 0.681 & 0.692 & 0.745 & 0.766 & 0.747 & 0.759 & 0.815 & 0.842 & 0.821 & 0.834 & 0.895 \\
\hline
HetGNN    & 0.706 & 0.690 & 0.701 & 0.760 & 0.774 & 0.756 & 0.767 & 0.840 & 0.868 & 0.847 & 0.862 & 0.915 \\
\hline
TGN       & 0.713 & 0.702 & 0.709 & 0.790 & 0.790 & 0.772 & 0.780 & 0.865 & 0.875 & 0.854 & 0.868 & 0.935 \\
\hline
DS-DGA-GCN & 0.715 & 0.699 & 0.708 & 0.805 & 0.784 & 0.766 & 0.776 & 0.865 & 0.883 & 0.858 & 0.866 & 0.928 \\
\hline
\end{tabularx}
\end{table*}

\subsubsection{Cross-Platform Adaptability Evaluation}

In addition to testing adaptability under temporal sparsity with newly launched products, 
we further investigate whether DS-DGA-GCN can generalize across different platforms. 
This setting evaluates robustness under domain shifts in user populations, interaction structures, and even language. 
We design two complementary transfer experiments: (i) Amazon$\leftrightarrow$Yelp, where both platforms share English-language reviews but differ in community norms and interaction patterns, thereby isolating platform-level transferability; and (ii) Amazon$\rightarrow$Xiaohongshu, which introduces both platform and cross-lingual shift (English reviews versus Chinese social posts).

\begin{table}[!t]
\caption{Cross platform transfer between Amazon and Yelp}
\label{tab:xfer_amz_yelp}
\centering
\setlength{\tabcolsep}{4pt} 
\renewcommand{\arraystretch}{1.2} 
\newcolumntype{C}{>{\centering\arraybackslash}X} 
\begin{tabularx}{\columnwidth}{|C|C|C|C|C|C|} 
\hline
\textbf{Train\newline$\rightarrow$Test} & \textbf{Setting} & \textbf{Accuracy} & \textbf{Recall} & \textbf{F1} & \textbf{AUROC} \\
\hline
Amazon\newline$\rightarrow$Yelp & Zero shot & 0.824 & 0.768 & 0.784 & 0.858 \\
\hline
Amazon\newline$\rightarrow$Yelp & +5\% target fine tuning & 0.852 & 0.802 & 0.821 & 0.898 \\
\hline
Yelp\newline$\rightarrow$Amazon & Zero shot & 0.834 & 0.780 & 0.796 & 0.864 \\
\hline
Yelp\newline$\rightarrow$Amazon & +5\% target fine tuning & 0.861 & 0.812 & 0.832 & 0.904 \\
\hline
\end{tabularx}
\end{table}

\begin{table}[!t]
\caption{Supplementary Amazon$\rightarrow$Xiaohongshu transfer under language and platform shift}
\label{tab:amz_to_xhs_shift}
\centering
\setlength{\tabcolsep}{4pt} 
\renewcommand{\arraystretch}{1.2} 
\newcolumntype{C}{>{\centering\arraybackslash}X} 
\begin{tabularx}{\columnwidth}{|C|C|C|C|C|} 
\hline
\textbf{Variant} & \textbf{Accuracy} & \textbf{Recall} & \textbf{F1} & \textbf{AUROC} \\
\hline
Structure only zero shot & 0.696 & 0.642 & 0.658 & 0.760 \\
\hline
Multilingual text embeddings zero shot & 0.714 & 0.658 & 0.676 & 0.792 \\
\hline
+5\% target fine tuning & 0.748 & 0.700 & 0.724 & 0.842 \\
\hline
\end{tabularx}
\end{table}

We first evaluate cross-platform adaptability between Amazon and Yelp, two English-language platforms with distinct interaction dynamics but without linguistic mismatch. This setting isolates platform-level domain shifts and provides a clearer measure of transferability. As shown in Table~\ref{tab:xfer_amz_yelp}, DS-DGA-GCN achieves reasonable zero-shot performance in both directions, with F1 scores around 0.78--0.80 and AUROC above 0.85. When only 5\% of labeled target data is available for fine-tuning, performance further improves by approximately 3--4 points across all metrics. These results indicate that DS-DGA-GCN successfully captures structural and temporal patterns that generalize across different review communities, demonstrating strong adaptability under realistic platform migration scenarios.

We also conduct a supplementary transfer from Amazon to Xiaohongshu, which introduces both cross-platform and cross-lingual shift (English reviews versus Chinese social posts). Results in Table~\ref{tab:amz_to_xhs_shift} show that the structure-only variant which incorporates the graph topology and our proposed NFS scores while excluding all textual embeddings yields modest transfer performance. The limited performance in this setting, despite the inclusion of NFS to capture platform-invariant structural anomalies, is primarily due to the extreme distribution shift of the initial node features ($H^u, H^p$). The lack of aligned semantic information between English and Chinese prevents the dynamic attention mechanism from accurately weighing interactions across such disparate feature spaces. While multilingual text embeddings provide incremental improvements and light fine-tuning with 5\% labeled target data further enhances results, the overall accuracy and F1 remain lower than in same-language transfer. This performance gap reflects the compounded difficulty of linguistic differences, heterogeneous content formats, and platform-specific behaviors. We conclude that DS-DGA-GCN retains partial adaptability under such extreme shifts, but its effectiveness is most clearly demonstrated in same-language transfer scenarios where platform-level generalization can be isolated and verified.

\section{Conclusion}
This paper proposes a novel framework, DS-DGA-GCN, for the adaptive detection of fake reviewer groups in dynamic networks. The core of the method involves using a NFS module to first quantify key structural features (e.g., self-similarity) into a node's prior anomaly score, which then guides a dynamic graph attention network to capture complex temporal and interaction patterns. Experimental results on real-world datasets like Amazon show that the model achieves excellent performance with up to 89.8\% accuracy, significantly outperforming existing baselines and demonstrating robust adaptability in data-sparse, cold-start scenarios. In summary, this research validates an effective design approach—combining interpretable structural features with dynamic graph learning models—which offers a valuable solution for tackling the challenges of online fraud detection in complex and evolving scenarios.



\bibliographystyle{IEEEtran}
\bibliography{ref}

@article{gai31kipf2016semi,
  title={Semi-supervised classification with graph convolutional networks},
  author={Kipf, Thomas N and Welling, Max},
  journal={arXiv preprint arXiv:1609.02907},
  year={2016}
}

@article{gai32velickovic2017graph,
  title={Graph attention networks},
  author={Velickovic, Petar and Cucurull, Guillem and Casanova, Arantxa and Romero, Adriana and Lio, Pietro and Bengio, Yoshua and others},
  journal={stat},
  volume={1050},
  number={20},
  pages={10--48550},
  year={2017}
}

@article{gai33vaswani2017attention,
  title={Attention is all you need},
  author={Vaswani, A},
  journal={Advances in Neural Information Processing Systems},
  year={2017}
}

@article{gai34zhang2018gaan,
  title={Gaan: Gated attention networks for learning on large and spatiotemporal graphs},
  author={Zhang, Jiani and Shi, Xingjian and Xie, Junyuan and Ma, Hao and King, Irwin and Yeung, Dit-Yan},
  journal={arXiv preprint arXiv:1803.07294},
  year={2018}
}

@article{gai38yu2017spatio,
  title={Spatio-temporal graph convolutional networks: A deep learning framework for traffic forecasting},
  author={Yu, Bing and Yin, Haoteng and Zhu, Zhanxing},
  journal={arXiv preprint arXiv:1709.04875},
  year={2017}
}

@article{gai21page1998pagerank,
  title={The pagerank citation ranking: Bringing order to the web. Technical report},
  author={Page, Lawrence},
  journal={Stanford Digital Library Technologies Project, 1998},
  year={1998}
}

@article{gai23akoglu2015graph,
  title={Graph based anomaly detection and description: a survey},
  author={Akoglu, Leman and Tong, Hanghang and Koutra, Danai},
  journal={Data mining and knowledge discovery},
  volume={29},
  pages={626--688},
  year={2015},
  publisher={Springer}
}

@article{gai24hamilton2017inductive,
  title={Inductive representation learning on large graphs},
  author={Hamilton, Will and Ying, Zhitao and Leskovec, Jure},
  journal={Advances in neural information processing systems},
  volume={30},
  year={2017}
}

@article{gai25kipf2016semi,
  title={Semi-supervised classification with graph convolutional networks},
  author={Kipf, Thomas N and Welling, Max},
  journal={arXiv preprint arXiv:1609.02907},
  year={2016}
}

@article{gai26velickovic2017graph,
  title={Graph attention networks},
  author={Velickovic, Petar and Cucurull, Guillem and Casanova, Arantxa and Romero, Adriana and Lio, Pietro and Bengio, Yoshua and others},
  journal={stat},
  volume={1050},
  number={20},
  pages={10--48550},
  year={2017}
}

@article{xiu1xu2020detect,
  title={Detect professional malicious user with metric learning in recommender systems},
  author={Xu, Yuanbo and Yang, Yongjian and Wang, En and Zhuang, Fuzhen and Xiong, Hui},
  journal={IEEE Transactions on Knowledge and Data Engineering},
  volume={34},
  number={9},
  pages={4133--4146},
  year={2020},
  publisher={IEEE}
}

@inproceedings{xiu2li2021large,
  title={Large-scale fake click detection for e-commerce recommendation systems},
  author={Li, Jingdong and Li, Zhao and Huang, Jiaming and Zhang, Ji and Wang, Xiaoling and Lu, Xingjian and Zhou, Jingren},
  booktitle={2021 IEEE 37th International Conference on Data Engineering (ICDE)},
  pages={2595--2606},
  year={2021},
  organization={IEEE}
}

@article{xiu3yang2019mining,
  title={Mining fraudsters and fraudulent strategies in large-scale mobile social networks},
  author={Yang, Yang and Xu, Yuhong and Sun, Yizhou and Dong, Yuxiao and Wu, Fei and Zhuang, Yueting},
  journal={IEEE Transactions on Knowledge and Data Engineering},
  volume={33},
  number={1},
  pages={169--179},
  year={2019},
  publisher={IEEE}
}

@article{xiu4fan2023adversarial,
  title={Adversarial attacks for black-box recommender systems via copying transferable cross-domain user profiles},
  author={Fan, Wenqi and Zhao, Xiangyu and Li, Qing and Derr, Tyler and Ma, Yao and Liu, Hui and Wang, Jianping and Tang, Jiliang},
  journal={IEEE Transactions on Knowledge and Data Engineering},
  volume={35},
  number={12},
  pages={12415--12429},
  year={2023},
  publisher={IEEE}
}

@inproceedings{xiu5hooi2016fraudar,
  title={Fraudar: Bounding graph fraud in the face of camouflage},
  author={Hooi, Bryan and Song, Hyun Ah and Beutel, Alex and Shah, Neil and Shin, Kijung and Faloutsos, Christos},
  booktitle={Proceedings of the 22nd ACM SIGKDD international conference on knowledge discovery and data mining},
  pages={895--904},
  year={2016}
}

@inproceedings{xiu6fayazi2015uncovering,
  title={Uncovering crowdsourced manipulation of online reviews},
  author={Fayazi, Amir and Lee, Kyumin and Caverlee, James and Squicciarini, Anna},
  booktitle={Proceedings of the 38th international ACM SIGIR conference on research and development in information retrieval},
  pages={233--242},
  year={2015}
}

@inproceedings{xiu7wang2019fdgars,
  title={Fdgars: Fraudster detection via graph convolutional networks in online app review system},
  author={Wang, Jianyu and Wen, Rui and Wu, Chunming and Huang, Yu and Xiong, Jian},
  booktitle={Companion proceedings of the 2019 World Wide Web conference},
  pages={310--316},
  year={2019}
}

@inproceedings{xiu8wang2021decoupling,
  title={Decoupling representation learning and classification for gnn-based anomaly detection},
  author={Wang, Yanling and Zhang, Jing and Guo, Shasha and Yin, Hongzhi and Li, Cuiping and Chen, Hong},
  booktitle={Proceedings of the 44th international ACM SIGIR conference on research and development in information retrieval},
  pages={1239--1248},
  year={2021}
}

@article{xiu9liu2018contrast,
  title={A contrast metric for fraud detection in rich graphs},
  author={Liu, Shenghua and Hooi, Bryan and Faloutsos, Christos},
  journal={IEEE Transactions on Knowledge and Data Engineering},
  volume={31},
  number={12},
  pages={2235--2248},
  year={2018},
  publisher={IEEE}
}

@article{xiu10gao2024revisiting,
  title={Revisiting Attack-caused Structural Distribution Shift in Graph Anomaly Detection},
  author={Gao, Yuan and Li, Jinghan and Wang, Xiang and He, Xiangnan and Feng, Huamin and Zhang, Yongdong},
  journal={IEEE Transactions on Knowledge and Data Engineering},
  year={2024},
  publisher={IEEE}
}

@article{xiu11yu2024temporal,
  title={Temporal Insights for Group-Based Fraud Detection on e-Commerce Platforms},
  author={Yu, Jianke and Wang, Hanchen and Wang, Xiaoyang and Li, Zhao and Qin, Lu and Zhang, Wenjie and Liao, Jian and Zhang, Ying and Yang, Bailin},
  journal={IEEE Transactions on Knowledge and Data Engineering},
  year={2024},
  publisher={IEEE}
}

@article{1111111he2022market,
  title={The market for fake reviews},
  author={He, Sherry and Hollenbeck, Brett and Proserpio, Davide},
  journal={Marketing Science},
  volume={41},
  number={5},
  pages={896--921},
  year={2022},
  publisher={INFORMS}
}

@article{2222222222tadelis2016reputation,
  title={Reputation and feedback systems in online platform markets},
  author={Tadelis, Steven},
  journal={Annual review of economics},
  volume={8},
  number={1},
  pages={321--340},
  year={2016},
  publisher={Annual Reviews}
}

@inproceedings{44444444mukherjee2012spotting,
  title={Spotting fake reviewer groups in consumer reviews},
  author={Mukherjee, Arjun and Liu, Bing and Glance, Natalie},
  booktitle={Proceedings of the 21st international conference on World Wide Web},
  pages={191--200},
  year={2012}
}

@article{5555555he2022detecting,
  title={Detecting fake-review buyers using network structure: Direct evidence from Amazon},
  author={He, Sherry and Hollenbeck, Brett and Overgoor, Gijs and Proserpio, Davide and Tosyali, Ali},
  journal={Proceedings of the National Academy of Sciences},
  volume={119},
  number={47},
  pages={e2211932119},
  year={2022},
  publisher={National Acad Sciences}
}

@inproceedings{777777lim2010detecting,
  title={Detecting product review spammers using rating behaviors},
  author={Lim, Ee-Peng and Nguyen, Viet-An and Jindal, Nitin and Liu, Bing and Lauw, Hady Wirawan},
  booktitle={Proceedings of the 19th ACM international conference on Information and knowledge management},
  pages={939--948},
  year={2010}
}

@inproceedings{888888olsson2024fakex,
  title={FakeX: A Framework for Detecting Fake Reviews of Browser Extensions},
  author={Olsson, Eric and Eriksson, Benjamin and Picazo-Sanchez, Pablo and Andersson, Lukas and Sabelfeld, Andrei},
  booktitle={Proceedings of the 19th ACM Asia Conference on Computer and Communications Security},
  pages={769--784},
  year={2024}
}

@inproceedings{9999999akoglu2013opinion,
  title={Opinion fraud detection in online reviews by network effects},
  author={Akoglu, Leman and Chandy, Rishi and Faloutsos, Christos},
  booktitle={Proceedings of the international AAAI conference on web and social media},
  volume={7},
  number={1},
  pages={2--11},
  year={2013}
}

@inproceedings{A11111111fayazi2015uncovering,
  title={Uncovering crowdsourced manipulation of online reviews},
  author={Fayazi, Amir and Lee, Kyumin and Caverlee, James and Squicciarini, Anna},
  booktitle={Proceedings of the 38th international ACM SIGIR conference on research and development in information retrieval},
  pages={233--242},
  year={2015}
}

@inproceedings{A22222wang2019fdgars,
  title={Fdgars: Fraudster detection via graph convolutional networks in online app review system},
  author={Wang, Jianyu and Wen, Rui and Wu, Chunming and Huang, Yu and Xiong, Jian},
  booktitle={Companion proceedings of the 2019 World Wide Web conference},
  pages={310--316},
  year={2019}
}

@inproceedings{A33333wang2021decoupling,
  title={Decoupling representation learning and classification for gnn-based anomaly detection},
  author={Wang, Yanling and Zhang, Jing and Guo, Shasha and Yin, Hongzhi and Li, Cuiping and Chen, Hong},
  booktitle={Proceedings of the 44th international ACM SIGIR conference on research and development in information retrieval},
  pages={1239--1248},
  year={2021}
}

@article{A444444zhao2023rhgnn,
  title={RHGNN: Fake reviewer detection based on reinforced heterogeneous graph neural networks},
  author={Zhao, Jun and Shao, Minglai and Tang, Hailiang and Liu, Jianchao and Du, Lin and Wang, Hong},
  journal={Knowledge-Based Systems},
  volume={280},
  pages={111029},
  year={2023},
  publisher={Elsevier}
}

@article{tuchihuaying2018hierarchical,
  title={Hierarchical graph representation learning with differentiable pooling},
  author={Ying, Zhitao and You, Jiaxuan and Morris, Christopher and Ren, Xiang and Hamilton, Will and Leskovec, Jure},
  journal={Advances in neural information processing systems},
  volume={31},
  year={2018}
}

@article{jinrui2023massive99,
  title={Massive Fake Review Group Recognition Based on Network Structure Features},
  author={Jinrui, WEI and Ruotong, WANG and Han, WANG},
  journal={Operations Research and Management Science},
  volume={32},
  number={1},
  pages={194},
  year={2023}
}

@inproceedings{soni2018effective1100,
  title={Effective machine learning approach to detect groups of fake reviewers},
  author={Soni, Jayesh and Prabakar, Nagarajan},
  booktitle={Proceedings of the 14th international conference on data science (ICDATA’18), Las Vegas, NV},
  pages={74--78},
  year={2018}
}

@article{gai99es2021mapreduce,
  title={A MapReduce opinion mining for COVID-19-related tweets classification using enhanced ID3 decision tree classifier},
  author={Es-Sabery, Fatima and Es-Sabery, Khadija and Qadir, Junaid and Sainz-De-Abajo, Beatriz and Hair, Abdellatif and Garcia-Zapirain, Begona and De La Torre-D{\'\i}ez, Isabel},
  journal={IEEE Access},
  volume={9},
  pages={58706--58739},
  year={2021},
  publisher={IEEE}
}

@article{li2021detectionAA12,
  title={Detection of fake reviews using group model},
  author={Li, Yuejun and Wang, Fangxin and Zhang, Shuwu and Niu, Xiaofei},
  journal={Mobile Networks and Applications},
  volume={26},
  pages={91--103},
  year={2021},
  publisher={Springer}
}

@inproceedings{cao2021fakeAA13,
  title={Fake reviewer group detection in online review systems},
  author={Cao, Chen and Li, Shihao and Yu, Shuo and Chen, Zhikui},
  booktitle={2021 International Conference on Data Mining Workshops (ICDMW)},
  pages={935--942},
  year={2021},
  organization={IEEE}
}

@article{song2005selfBB24,
  title={Self-similarity of complex networks},
  author={Song, Chaoming and Havlin, Shlomo and Makse, Hernan A},
  journal={Nature},
  volume={433},
  number={7024},
  pages={392--395},
  year={2005},
  publisher={Nature Publishing Group UK London}
}

@article{gallos2007reviewBB25,
  title={A review of fractality and self-similarity in complex networks},
  author={Gallos, Lazaros K and Song, Chaoming and Makse, Hern{\'a}n A},
  journal={Physica A: Statistical Mechanics and its Applications},
  volume={386},
  number={2},
  pages={686--691},
  year={2007},
  publisher={Elsevier}
}

@article{song2005selfBB29,
  title={Self-similarity of complex networks},
  author={Song, Chaoming and Havlin, Shlomo and Makse, Hernan A},
  journal={Nature},
  volume={433},
  number={7024},
  pages={392--395},
  year={2005},
  publisher={Nature Publishing Group UK London}
}

@inproceedings{wang2012recommendationCC32,
  title={Recommendation based on mining product reviewers preference similarity network},
  author={Wang, Feng and Chen, Li},
  booktitle={Proceedings of 6th SNAKDD Workshop},
  pages={166},
  year={2012}
}

@article{palla2008fundamentalCC34,
  title={Fundamental statistical features and self-similar properties of tagged networks},
  author={Palla, Gergely and Farkas, Ill{\'e}s J and Pollner, P{\'e}ter and Der{\'e}nyi, Imre and Vicsek, Tam{\'a}s},
  journal={New Journal of Physics},
  volume={10},
  number={12},
  pages={123026},
  year={2008},
  publisher={IOP Publishing}
}

@book{kelley2017generalCC37,
  title={General topology},
  author={Kelley, John L},
  year={2017},
  publisher={Courier Dover Publications}
}

@article{zhou2022networkCC38,
  title={Network representation learning: from preprocessing, feature extraction to node embedding},
  author={Zhou, Jingya and Liu, Ling and Wei, Wenqi and Fan, Jianxi},
  journal={ACM Computing Surveys (CSUR)},
  volume={55},
  number={2},
  pages={1--35},
  year={2022},
  publisher={ACM New York, NY}
}

@article{xu2020detect,
  title={Detect professional malicious user with metric learning in recommender systems},
  author={Xu, Yuanbo and Yang, Yongjian and Wang, En and Zhuang, Fuzhen and Xiong, Hui},
  journal={IEEE Transactions on Knowledge and Data Engineering},
  volume={34},
  number={9},
  pages={4133--4146},
  year={2020},
  publisher={IEEE}
}

@article{si2020shilling,
  title={Shilling attacks against collaborative recommender systems: a review},
  author={Si, Mingdan and Li, Qingshan},
  journal={Artificial Intelligence Review},
  volume={53},
  pages={291--319},
  year={2020},
  publisher={Springer}
}

@inproceedings{mukherjee2013yelp,
  title={What yelp fake review filter might be doing?},
  author={Mukherjee, Arjun and Venkataraman, Vivek and Liu, Bing and Glance, Natalie},
  booktitle={Proceedings of the international AAAI conference on web and social media},
  volume={7},
  number={1},
  pages={409--418},
  year={2013}
}

@article{yu2024temporal,
  title={Temporal Insights for Group-Based Fraud Detection on e-Commerce Platforms},
  author={Yu, Jianke and Wang, Hanchen and Wang, Xiaoyang and Li, Zhao and Qin, Lu and Zhang, Wenjie and Liao, Jian and Zhang, Ying and Yang, Bailin},
  journal={IEEE Transactions on Knowledge and Data Engineering},
  year={2024},
  publisher={IEEE}
}

@article{liu2018contrast,
  title={A contrast metric for fraud detection in rich graphs},
  author={Liu, Shenghua and Hooi, Bryan and Faloutsos, Christos},
  journal={IEEE Transactions on Knowledge and Data Engineering},
  volume={31},
  number={12},
  pages={2235--2248},
  year={2018},
  publisher={IEEE}
}

@article{velivckovic2017graph,
  title={Graph attention networks},
  author={Veli{\v{c}}kovi{\'c}, Petar and Cucurull, Guillem and Casanova, Arantxa and Romero, Adriana and Lio, Pietro and Bengio, Yoshua},
  journal={arXiv preprint arXiv:1710.10903},
  year={2017}
}

@article{wu2020fake,
  title={Fake online reviews: Literature review, synthesis, and directions for future research},
  author={Wu, Yuanyuan and Ngai, Eric WT and Wu, Pengkun and Wu, Chong},
  journal={Decision Support Systems},
  volume={132},
  pages={113280},
  year={2020},
  publisher={Elsevier}
}

@inproceedings{new11zhang2019heterogeneous,
  title={Heterogeneous graph neural network},
  author={Zhang, Chuxu and Song, Dongjin and Huang, Chao and Swami, Ananthram and Chawla, Nitesh V},
  booktitle={Proceedings of the 25th ACM SIGKDD international conference on knowledge discovery \& data mining},
  pages={793--803},
  year={2019}
}

@article{new12rossi2020temporal,
  title={Temporal graph networks for deep learning on dynamic graphs},
  author={Rossi, Emanuele and Chamberlain, Ben and Frasca, Fabrizio and Eynard, Davide and Monti, Federico and Bronstein, Michael},
  journal={arXiv preprint arXiv:2006.10637},
  year={2020}
}

@article{new13youden1950index,
  author  = {Youden, W. J.},
  title   = {Index for Rating Diagnostic Tests},
  journal = {Cancer},
  year    = {1950},
  volume  = {3},
  number  = {1},
  pages   = {32--35}
}

@article{new211fawcett2006introduction,
  title={An introduction to ROC analysis},
  author={Fawcett, Tom},
  journal={Pattern recognition letters},
  volume={27},
  number={8},
  pages={861--874},
  year={2006},
  publisher={Elsevier}
}

@inproceedings{new333mukherjee2012spotting,
  title={Spotting fake reviewer groups in consumer reviews},
  author={Mukherjee, Arjun and Liu, Bing and Glance, Natalie},
  booktitle={Proceedings of the 21st international conference on World Wide Web},
  pages={191--200},
  year={2012}
}

\begin{IEEEbiography}[{\includegraphics[width=1in,height=1.25in,clip,keepaspectratio]{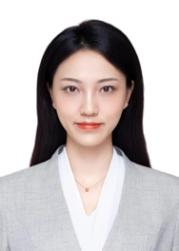}}]{Jing Zhang} received her Ph.D. degree in computer science and technology from Northwest University, Xi'an, China, in 2018. She is an Associate Professor with Xi’an University of Science and Technology, Xi’an. She has been presided two NSFC projects and one project from Shaanxi Science and Technology Department. Her research interests include ubiquitous computing, mobile crowd sensing and human-AI collaboration.
\end{IEEEbiography}

\begin{IEEEbiography}
[{\includegraphics[width=1in,height=1.25in,clip,keepaspectratio]{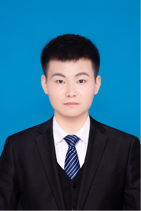}}]{Ke Huang} is currently a Ph.D. student at the School of Computer Science, Northwestern Polytechnical University. His current research interests include fraud detection, dynamic network analysis, and graph representation learning. He has published several research papers in international journals and conferences, including IEEE Internet of Things Journal.
\end{IEEEbiography}

\begin{IEEEbiography}
[{\includegraphics[width=1in,height=1.25in,clip,keepaspectratio]{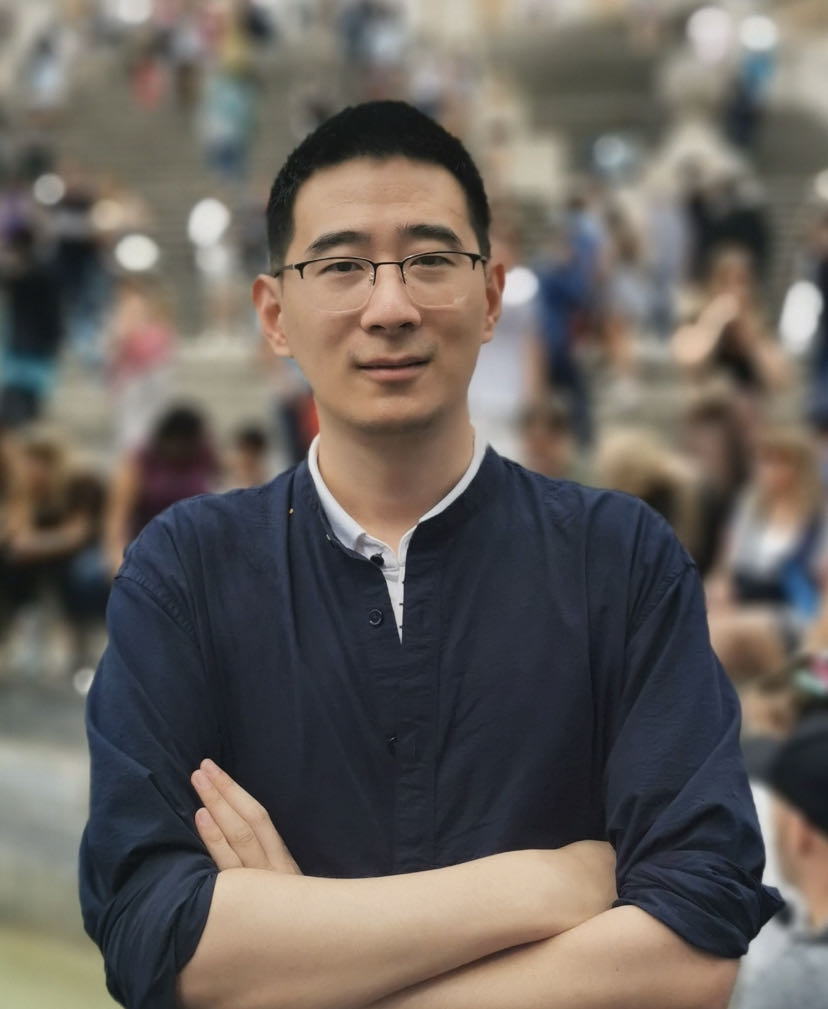}}]{Yao Zhang} received his Ph.D. degree in communication and information system from Xidian
University, Xi'an, China, in 2020. He served as a Research Assistant and Post-Doctoral Fellow at The Hong Kong Polytechnic University in 2019 and 2021, respectively. From 2021 to 2023, he
was a Post-Doctoral Researcher at Northwestern Polytechnical University, Xi'an, China, where he is currently an Associate Professor. His research interests include CrowdNet, human-AI
collaboration, mobile edge computing, edge AI, and networked autonomous driving.
\end{IEEEbiography}

\begin{IEEEbiography}
[{\includegraphics[width=1in,height=1.25in,clip,keepaspectratio]{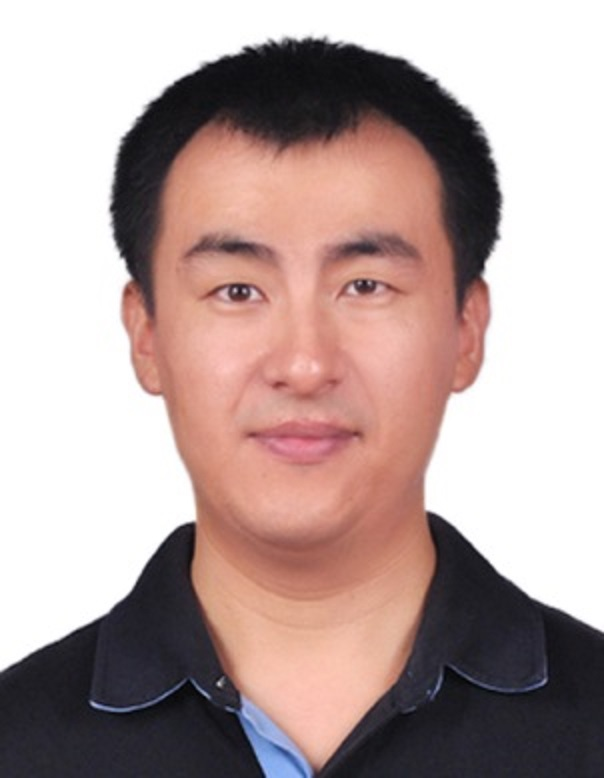}}]{Bin Guo} received the PhD degree in computer science from Keio University, Minato, Japan, in 2009, and then was a postdoc researcher with Insti-
tut Telecom SudParis, France. He is currently a professor with Northwestern Polytechnical University, Xi'an, China. His research interests include ubiquitous computing, mobile crowd sensing, and human-computer interaction. He has served as an associate editor of the IEEE Communications Magazine and the IEEE Transactions on Human-Machine-Systems, the guest editor of the ACM Transactions on Intelligent Systems and Technology and the IEEE Internet of Things, the general co-chair of IEEE UIC'15, and the program chair of IEEE CPSCom'16, ANT'14, and UIC'13. He is a senior member of the IEEE.
\end{IEEEbiography}

\begin{IEEEbiography}
[{\includegraphics[width=1in,height=1.25in,clip,keepaspectratio]{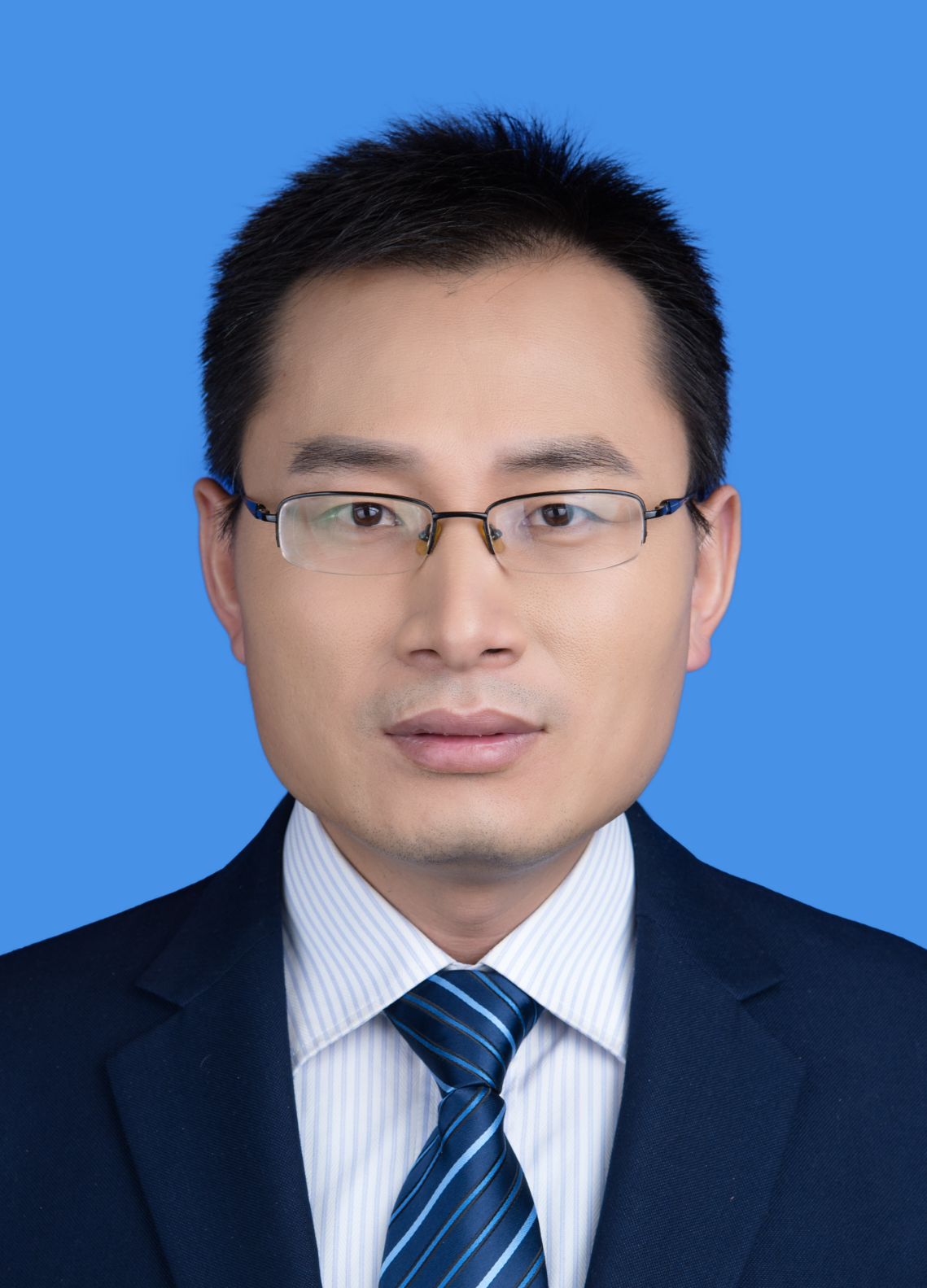}}]{Zhiwen Yu} 
received the Ph.D. degree of engineering in computer science and technology from Northwestern Polytechnical University, Xi'an, China, in 2005. He is currently with Harbin Engineering University, Harbin, Heilongjiang, China, and also a professor at Northwestern Polytechnical University, Xi'an, China. He has worked as a research fellow with the Academic Center for Computing and Media Studies, Kyoto University, Japan, from February 2007 to January 2009, and a post-doctoral researcher with the Information Technology Center, Nagoya University, Japan, in 2006-2007. His research interests include pervasive computing, context-aware systems, human-computer interaction, mobile social networks, and personalization. He is the associate editor or editorial board member for the IEEE Communications Magazine, the IEEE Transactions on Human-Machine Systems, Personal and Ubiquitous Computing, and Entertainment Computing (Elsevier). He was the general chair of IEEE CPSCom'15, and IEEE UIC'14. He served as a vice program chair of PerCom'15, the program chair of UIC'13, and the workshop chair of UbiComp'11. He is a senior member of the IEEE, a member of the ACM, and a council member of the China Computer Federation (CCF).
\end{IEEEbiography}

\vfill

\end{document}